\documentclass[prb,twocolumn,nofootinbib]{revtex4-1}

\usepackage[ruled,vlined]{algorithm2e}

\usepackage{graphicx}  
\usepackage{subfigure}
\usepackage{multirow}

\usepackage{fancyhdr}
\usepackage{longtable}
\usepackage{parskip}
\usepackage[T1]{fontenc}
\usepackage{dcolumn}   
\usepackage{mathtools}

\usepackage{bm}        
\usepackage{amsfonts}  
\usepackage{amsmath}   
\usepackage{amssymb}   

\graphicspath{{figures/}}

\begin{document}

\title{Matrix product state recursion methods for computing spectral functions of strongly correlated quantum systems}
\author{Yifan Tian and Steven R. White}
\affiliation{Department of Physics and Astronomy, University of California, Irvine CA 92697, USA
s}
\date{\today}

\begin{abstract}
We present a method for extrapolation of real-time dynamical correlation functions which can improve the capability of matrix product state methods to compute spectral functions. Unlike the widely used linear prediction method, which ignores the origin of the data being extrapolated, our recursion methods utilize a representation of the wavefunction in terms of an expansion of the same wavefunction and its translations at earlier times. This recursion method is exact for a noninteracting Fermi system. Surprisingly, the recursion method is also more robust than linear prediction at large interaction strength.  We test this method on the Hubbard two-leg ladder, and present more accurate results for the spectral function than previous studies. 
\end{abstract}

\maketitle

\section{\label{sec:one}Introduction}


Time-dependent matrix product state (MPS) methods are widely used to compute spectral function of strongly interacting systems in one dimension, and increasingly in two dimensions.  These methods, however, have a fundamental limitation: the entanglement of the wavefunction grows with the time.  Ground states are governed by the area law, and thus have low entanglement, particularly in one dimension. MPS representations of ground states are readily obtained with the density matrix renormalization group (DMRG)\cite{Steven_1992,Steven_1993}.  
When the ground state is altered by, say, adding a particle or flipping
a spin at one site, the entanglement is unchanged. However, as time evolves, the local change in the wavefunction spreads through the system, causing growth of entanglement. The increased entanglement destroys the compression of the MPS representation and limits the total time for which accurate correlations can be obtained.  

Fortunately, the growth of entanglement induced by a local change in the wavefunction (a local quench) is much smaller than a global quench\cite{Quench_Dynamics, Quantum_quench}.(ADD Ref)  
The frequency resolution of the final spectrum is inversely related to the simulation time.
For 1D systems, one can typically continue the MPS simulations long enough  to get 
good results, but one would always like to have more time data.  

To make the best use of the data one has, extrapolation methods which approximately extend the data to larger times are very useful. This is because the Fourier transform to go from a time dependent function to frequency space requires a windowing function, smoothly reducing the signal to zero at the edges of the time range, in order to avoid unphysical oscillations. At the edges of the time window, the data does not have to be very accurate, since it is multiplied by the small value of the window function. The linear prediction method  (LP) serves as a general purpose extrapolator for time series of this sort.\cite{Numerical_Recipes} It has been used to extrapolate nuclear magnetic resonance signals since the 1980s \cite{STEPHENSON1988515}. It was first use in an MPS context by one of the authors\cite{Steven_2008}. Linear prediction is roughly equivalent to fitting the existing data with a set of decaying sine waves, and following the fit. In typical spectral function calculations this extrapolation is very accurate at first but it gradually loses accuracy.  One can extend one's data up to the point where unphysical artifacts appear in the spectral functions; typically, this is two to three times the length of the original data.

Any extrapolation method which could extend the useful time range further would be highly useful. Here we discuss an alternative extrapolation formulation based on the fact that there is an underlying wavefunction being evolved, rather than simply a time series of data. The recursion methods we describe are based on the idea that the wavefunction at one time can be approximated in terms of the wavefunction at earlier times. Since from our simulation we already know how the earlier time wavefunctions have evolved, we can evolve beyond the range of the simulation data
using the expansion.  By repeatedly using the expansion, one can extrapolate to infinite times.  A key part of our recursion methods is that in our expansion-set of earlier wavefunctions we include all possible translations of the wavefunction.  With the translations, recursion becomes exact in the case of a single particle, and also for a noninteracting set of fermions.  One also expects it to be very accurate when a single quasiparticle peak dominates the spectrum for all momenta.

We further generalize a recursion at one time to a multi-recursion method, where the recursion process is  repeated, say five or ten times.  In testing this method on the two-leg Hubbard ladder, we find that, surprisingly,  multi-recursion works well even in the large-$U$ regime.  Generally, it seems to perform better than linear prediction, at least for this model. We also note that there are a number of variations in how one performs the recursion, so we expect further improvements in our methods as we gain more experience. 

There are already a number of alternative methods for obtaining spectral functions with MPS methods.  A number of these do not involve time evolution.  Some methods work directly in frequency space, such as the correction vector method\cite{Correction_Vector1,Correction_Vector2} and its improvement, dynamical DMRG\cite{Jeckelmann2002,Jeckelmann2004}. Other methods work with powers of the Hamiltonian operator, such as the Chebyshev method\cite{PhysRevB.83.195115, PhysRevB.91.115144}.   Which method
is the best choice may depend on the particular problem being studied, or even on computational details, such as whether one has a cluster with many nodes available. (The frequency methods are perfectly parallelizable over frequencies, since each frequency point is a separate calculation; in time dependent methods, there is only one (longer) calculation, followed by Fourier transforms.) Here, we will restrict our focus to the time dependent methods, in particular those where currently one would use
linear prediction to extend the data, and evaluate how our recursion methods compare with linear prediction. 

We present results for the density of states and spectral functions for the 2-leg Hubbard ladder as a function of the interaction $U$ and the doping. This system is much more challenging than, say, a simple spin chain, so our results serve as a demonstration of what we can currently achieve with current MPS methods coupled with recursion. This spectral functions of Hubbard models have  been much studied in the past, including on ladders, but much of it was done quite some time ago, and even fairly recent studies do not necessarily reflect the current state of the art.   The half-filled ladder was studied using quantum Monte Carlo and maximum entropy methods two decades ago.\cite{Endres1996} More recent work was based on  tDMRG\cite{Feiguin2019}, and dynamical DMRG\cite{Masanori2020}. Also related is a Lanczos study of the $t$-$J$ ladder\cite{Martins1999}. An accurate study of a single-hole doped $t$-$J$ ladder has been performed with tDMRG techniques\cite{White2015}.

In the next section, we will derive and discuss the recursion methods. In Section III, we will review the time evolution methods, briefly describing our choice here, the time dependent variational principle(TDVP)\cite{TDVP}. We will also review linear prediction. In Section IV, we will test the recursion methods and compare them with linear prediction, taking as test systems a Hubbard chain and a two-leg Hubbard ladder. In Section V, we use recursion as part of a high-resolution study of the spectral function of the two-leg ladder as a function of $U/t$ and for several dopings.  Finally, in Section VI, we conclude. The software artifacts associated with this paper's methods and results are available at GitHub: \url{https://github.com/YifanTian/MPS_recursion_method}.

\section{\label{sec:three}Recursion methods}
The positive frequency part of the ground state spectral function for a spinful fermion system, such as the Hubbard model, can be obtained from the time evolution of the  wave-function  $|\psi_{0}(t)\rangle = e^{-iHt} c_{0,\uparrow}^{\dagger} |0\rangle$. (Similarly, the negative frequency part is obtained by replacing $c_{0,\uparrow}^\dagger$ by $c_{0,\uparrow}$, and similarly look at down-spins.) Here the origin $0$ indicates the site at which the particle is added, which we take to be at or near the center of a long chain or ladder, and $|0\rangle$ is the ground state (at fixed fillings, half-filling is one particle per site). We define $|\psi_{i}(t)\rangle$ similarly, as the wavefunction obtained from adding the particle at site $i$. 

Consider first the case where the ground state is the vacuum.  During the time evolution, the added particle spreads out in a wavepacket, and its wavefunction at a particular time $t_1$ can be represented exactly in terms of the $|\psi_{i}(0)\rangle$ as 
\begin{equation} 
|\psi_0(t_1)\rangle = \sum_j C_{j} |\psi_j(0)\rangle.
\label{psi0}
\end{equation}
If we then apply a time evolution operator to both sides of Eq. (\ref{psi0}), we obtain
\begin{equation} 
|\psi_0(t)\rangle = \sum_j C_{j} |\psi_j(t-t_1)\rangle \ \ \  \  t\ge t_1
\label{psi0t}
\end{equation}
Next, consider a noninteracting Fermi system, where the ground state is a filled Fermi sea.  Now suppressing the spin index,  $c_0^\dagger$ can be decomposed into operators which create single particle eigenstates, call them $c_k^\dagger$.  The time evolution induced by $c_k^\dagger$ acting on the ground state is trivial: either the ground state is destroyed if the state is occupied, or a new stationary state is created for which only the phase changes. In any case, these states can again be formed from the set of $c_i^\dagger$, so Eq. (\ref{psi0t}) again holds exactly. 

If translational invariance holds in a form allowing us to obtain $|\psi_j(t)\rangle$ from $|\psi_0(t)\rangle$, then in principle we could use the recursion relation Eq. (\ref{psi0t}) to find the wavefunction at any later time based on the original time evolution up to the time $t_1$. However, in practice the actual recursion is performed on the Green's function rather than the wavefunction; see below.

For a general interacting system we write for $t>t_1$
\begin{equation} 
|\psi_i(t)\rangle = \sum_j C_{ij} |\psi_j(t-t_1)\rangle + |R_i(t)\rangle.
\label{defineR}
\end{equation} 
Focusing on $i=0$, we choose the coefficients $C_{0j}$ to minimize the norm of the {\it residual wavefunction}  $|R(t)\rangle \equiv |R_0(t)\rangle$, which is then defined by this equation. In this case, we can recursively evolve the wavefunction by either neglecting or approximating $R$, or more practically, we can find the future Green's function by using recursion, neglecting or approximating the part of the Green's function due to $R$. 
 
Define a ``Green's function'' (without time ordering or factors of $i$) as
\begin{equation}
 G_{kj}(t) = \langle 0|c_k(t) c^\dagger_j(0)|0\rangle = \langle\psi_k(0)|\psi_j(t)\rangle.
\end{equation}
We call the equal-time Green's function the overlap matrix $O_{jk} = G_{jk}(0)$, since we treat the $\psi_j$ as basis functions. The norm squared of the residual is then
\begin{equation} 
\langle R|R\rangle = O_{0,0} - 2{\rm Re}\left[\sum_j C_{j}^* G_{j0}(t_1)\right] + \sum_{i,j} C_{i}^* O_{ij} C_{j} .
\end{equation} 
Minimizing the norm, we find 
\begin{equation}
C_{j} = [O^{-1}]_{jk}G_{k0}(t_1)
\end{equation}
For the time evolution started from adding a particle at $i$, we generalize this to $C_{ij} = [O^{-1}]_{jk}G_{ik}(t_0)$.

We see that the coefficients of the recursion of the wavefunction are defined in terms of the Green's function, which is what is needed for the spectral function. Applying $\langle \psi_k(0)|$ to Eq. (\ref{defineR}), we find
\begin{equation} 
G_{ki}(t) = \sum_j C_{ij} G_{kj}(t-t_1) + \langle\psi_k(0)|R_i(t)\rangle.
\label{Grecur}
\end{equation} 

The ``Green's function of the residual'' $R_{ki}(t) = \langle\psi_k(0)|R_i(t)\rangle$ has an important property:  it is zero at $t=t_1$, even when the residual is not small.  This is because $\langle\psi_k(0)$ is one of the basis functions used in defining the $C_{ij}$,
so the leftover part, the residual, has no component in that direction.  Thus, even neglecting the residual completely, the recursion gives the right answer for the Green's function very close to $t_1$. One might imagine that the residual could contain a large incoherent superposition of energy levels which have become out of phase with the wavefunction at time 0; the recursion would not capture these states well at all, but it wouldn't matter (near $t_1$) since these states would not contribute to the Green's function.

\begin{figure}[h]
\includegraphics[width=1.0\linewidth]{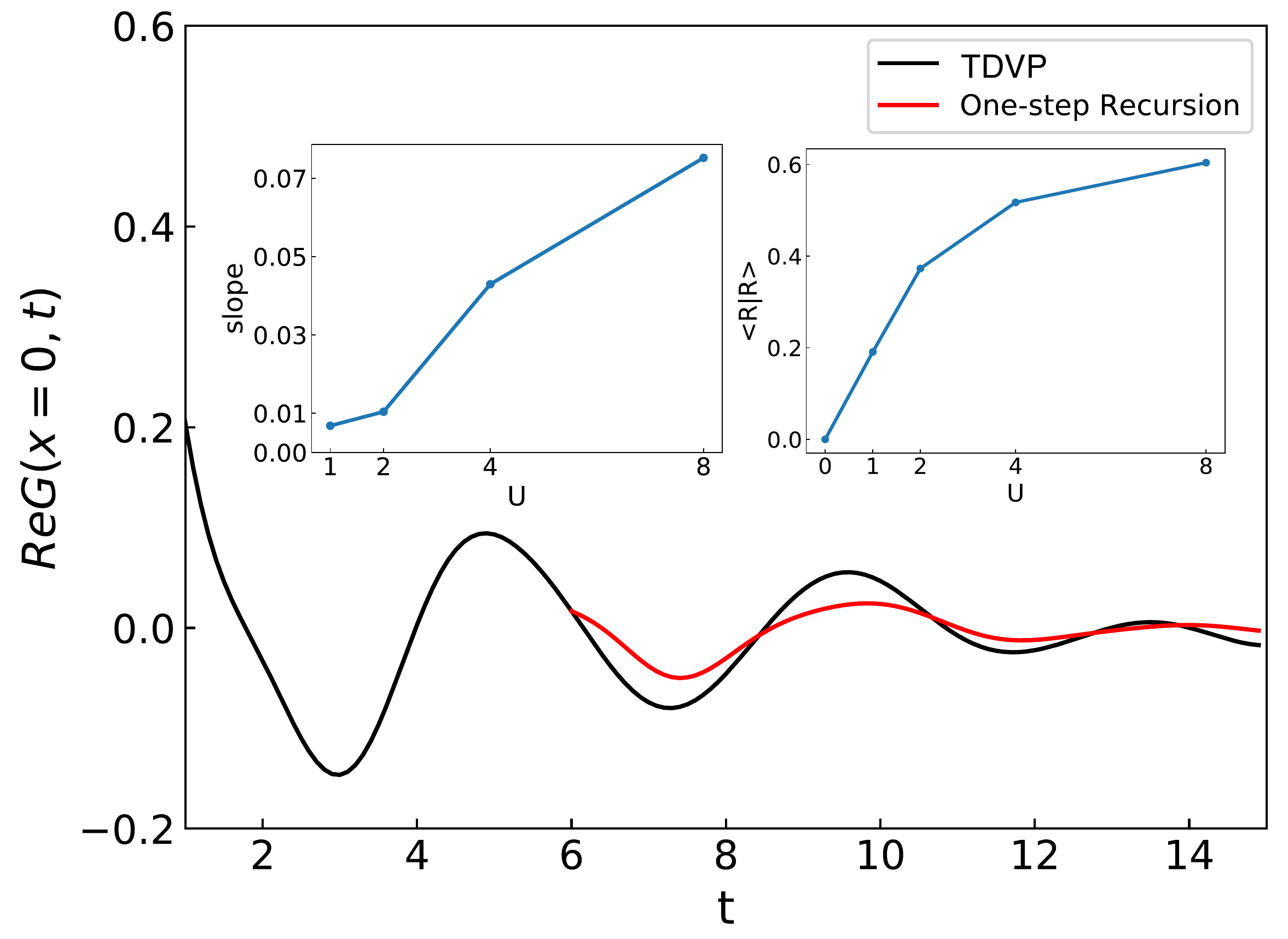}
\caption{The real part of the single particle Green's function $G(x=0,t)$ for the Hubbard ladder at U=8 on a $64\times2$ lattice at 1/8 hole doping. The black curve represents DMRG data out to t = 15; the red curve is a one-step recursion extrapolation for $t_1=6$, which is based on the data for $0 \le t \le t_1$.
The left inset shows the magnitude of the slope discontinuity after single-step projection versus $U$.
The right inset shows the norm squared of the residual $\langle R|R\rangle$ after single-step projection versus $U$, also at $p=1/8$.}
\label{fig:discontinuity}
\end{figure}

The simplest recursion is to neglect the second term in Eq. (\ref{Grecur}). We first use MPS-based time evolution to evolve the wavefunction and compute the Green's function up to time $t_1$. It is then straightforward to evolve the Green's function forward in time, time-step by time-step, with a negligible computational cost. We call this {\it one-step recursion}.  In Figure \ref{fig:discontinuity}, we compare the resulting extrapolated Green's function with the DMRG calculation performed past $t_1$ in the case of a 2-leg $2\times64$ Hubbard ladder at U=8 with 1/8 hole-doping.  (In practice, we would start the recursion at the last available DMRG data point.) We can see that the error in the recursion starts at 0 at $t_1$ but the slope of the error is non-zero, giving rise to a slope discontinuity. The magnitude of the slope discontinuity is shown in the left inset; it is zero at $U=0$ and rises roughly linearly with $U$. This slope discontinuity is the key limitation of the one-step recursion since it can give unphysical oscillations in the resulting spectrum. The right inset shows the magnitude of the residual after the recursion, which are quite sizeable. The errors in the extrapolation of the Green's function, however, are much smaller, but still too large to be very useful. 

There are several ways to go beyond the single-step recursion.  In the first method we consider, which we call {\it multi-step recursion},  
we use multiple projections to further reduce the norm of $|R\rangle$.  The first projection fits the original wavefunction, after which the second fits the residual of the first, etc.  All of the projections can be written in terms of the Green's functions, and we will use the term residual loosely to mean either the wavefunction residual or the associated Green's function. The first residual (for $t>t_1$) is 
\begin{equation} 
R^{1}_{ki}(t) = G_{ki}(t) - \sum_{j}  G_{kj}(t-t_1) C^{1}_{ji} .
\end{equation}
We project the residual piece at $t_2$ using coefficients
\begin{equation} 
C^{2}_{ji} = \sum_k [O^{-1}]_{jk}R^{1}_{ki}(t_2) .
\end{equation} 
After this projection, the second residual is
\begin{equation} 
R^{2}_{ki}(t) = R^{1}_{ki}(t) - \sum_{j}  G_{kj}(t-t_2)C^{2}_{ji}.
\end{equation}
After $M$ projections, we have
\begin{equation} 
R^{M}_{ki}(t) = R^{M-1}_{ki}(t)  - \alpha_{M}\sum_{j}  G_{kj}(t-t_M) C^{M}_{ji}.
\end{equation}
We now neglect $R^{M}$, setting it to zero, to form the extrapolation of the Green's function
\begin{equation}
G_{ki}(t) = \sum_{m=1}^{M} \sum_{j} G_{kj}(t-t_m) C^{m}_{ji},\ \  \ t>t_M
\label{multi_Re}
\end{equation}
This equation can be evolved time-step by time-step starting at $t_M$.

\begin{figure}[h]
\includegraphics[width=1.0\linewidth]{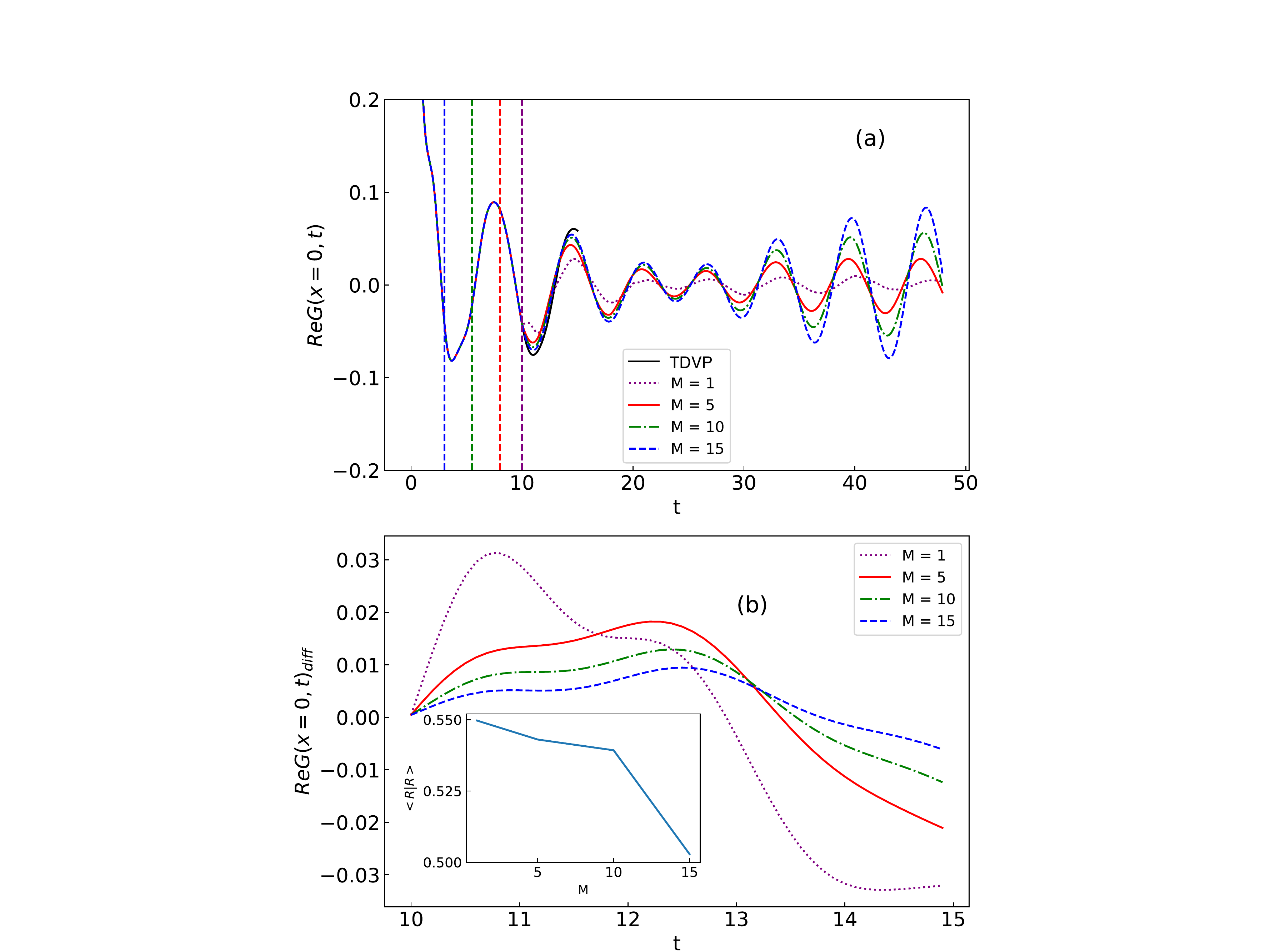}
\caption{(a) The long-term extrapolation of the Green's function using multi-step recursion for a 2-leg Hubbard ladder on a $64\times2$ lattice at U=8 with 1/4 hole-doping for t > 10 versus the number of projections $M$ with the fixed spacing of 0.5. Only the real part of the on-site Green's function is shown. An instability is evident in the extrapolations, particularly for $M=15$. The vertical lines shows where the first recursion happens for different multistep recursion. (b) The corresponding error in the Green's function at shorter times. The inset shows the $\langle R|R\rangle$ of each multistep recursion.}
\label{multi_projection_study}
\end{figure}
If we choose to perform the recursions at equally-spaced times, there are two adjustable parameters in multistep recursion, the number of recursions $M$ and the recursion interval $t_s$. (Except in testing, we make the last recursion at the end of the DMRG data). 
Several questions arise: does multistep recursion eliminate the slope discontinuity at $t_M$? Are the multistep extrapolations better than single step recursion? Does recursion have instabilities where the extrapolation grows without bound? What are the optimal choices for the two parameters? 

We find that multistep recursion performs significantly better than single-step recursion.  The slope discontinuity is not eliminated, but it becomes significantly smaller.  We find that instability can occur, particularly if a large number of recursions are performed, but the instability is not particularly bad if one only wants to extrapolate out, say, to about $4 t_M$ and apply a windowing function to the data, and also below we give a way to eliminate any instabilities. Typical results are shown in Fig~\ref{multi_projection_study}.  
For this model at $U=8$, a spacing around 0.5 is usually a good choice. Small spacing and many projections tends to  make the instability more of a problem.   Making $t_s$ larger leads to fewer projections and some loss of accuracy. A reasonable number of projections is $10-12$. More results and comparisons with linear predictions are shown in Section IV. 

A key issue in implementing recursion is how to use translational invariance. It may be possible to use an infinite MPS method to allow perfect translational invariance, but in our more conventional simulations we simulate on a long open chain or ladder. The overlap matrix $O_{ij}$ is the single particle equal time Green's function. We perform the simulation on a system of length $L$, which is large enough so that the signal induced by adding or removing a particle has not hit the edge of the system at the maximum simulation time.  In order to use recursion for longer times, the recursion should be done on a larger system to avoid having the extrapolated Green's function hit the edges at these longer times. (The recursion does not take a significant amount of computer time, so it is easy to use a long system.)

Calculating the full overlap matrix for all sites $i$ and $j$ are ideal, but we have found that fixing $i=0$ and using translational invariance  ($O_{ij} = O_{0,j-i}$ in 1D or its generalization to the ladder) is generally fine, particularly at large $U$, where the equal time Green's function decays more rapidly on the ladder. Instead of an inverse of $O$, we use a pseudo-inverse, which takes care of small or even slightly negative eigenvalues of the overlap due to the approximations made. A pseudo-inverse has been applied to stabilize the inverse of a matrix with small eigenvalues in previous time-evolution studies as well \cite{Spectral_functions_finite_T}.  As another optimization, we use reflection symmetry to reduce errors in calculating $O$, averaging results that should be the same by symmetry.

Instabilities like we see in recursion also occur in linear prediction. A standard part of using linear prediction is to remove the instabilities. We can use essentially the same technique to remove the instabilities in recursion. This underscores similarities in linear prediction and recursion, although the wavefunction nature of recursion opens up a variety of possible extensions outside the scope of linear prediction. In both methods, the technique to fix instabilities involves writing the extrapolation in terms of powers of a matrix.  One then
finds the complex eigenvalues with magnitude greater than 1,  reduces their magnitude to 1, and then reassembles the matrix.

To implement this, we first rewrite  Eq. ({\ref{multi_Re}}) in matrix language, where the site indices $i$, $j$, and $k$ are omitted as the matrix indices of blocks living in larger vectors
\begin{equation}
G(t)=
\left(\begin{array}{cccccc}
G_{t-t_1} & .. & G_{t-t_{M}} \\
\end{array} \right)
\left(\begin{array}{cccccc}
C^{1} \\ : \\ C^{M} \\
\end{array} \right)
\end{equation}
To allow the iteration to be written as powers of a matrix, we include $G(t)$ as part of a vector of different time $G's$, shifted by one time interval from the $G$-vector above.   To make this clear, we specialize to the case $M=3$ and write out the vectors and matrices completely:
\begin{equation}
\begin{multlined}
\left(\begin{array}{ccc}
G_{t} & G_{t-t_1} & G_{t-t_2}
\end{array} \right)
=
\\ 
\left(\begin{array}{ccc}
G_{t-t_{1}} & G_{t-t_{2}} & G_{t-t_{3}}  \\
\end{array} \right)
\left(\begin{array}{ccc}
C_{1} & I & 0  \\
C_{2} & 0 & I   \\
 C_{3} & 0 & 0  \\
\end{array} \right)
\end{multlined}
\label{Multi_Re_Matrix}
\end{equation}
Assuming that the recursion times $t_m$ are equally spaced, with $d$ time steps separating the recursion times, then the recursion splits into $d$ separate recursions.  For example, if $d=2$, the odd (even) time steps are used to predict future odd (even) time steps. Each sub-recursion is controlled by the same large matrix, i.e. the matrix in Eq. ({\ref{Multi_Re_Matrix}}) (for $M=3$). Now, to remove the instabilities, we diagonalize this matrix, and reduce the magnitude of any eigenvalues which are bigger than 1 to 1.
We then reassemble the matrix and use it for the recursion. In practice, we have found that typically the magnitude of a small number of eigenvalues is slightly large than 1, e.g. by $\sim 10\%$, and that this procedure is satisfactory in removing the instability. (One further note: in order to reduce errors from the diagonalization, we compute only the changes to the matrix induced by the altered eigenvalues, and add the changes to the original matrix.)

\begin{figure}[h!]
\centering
\includegraphics[scale=0.5]{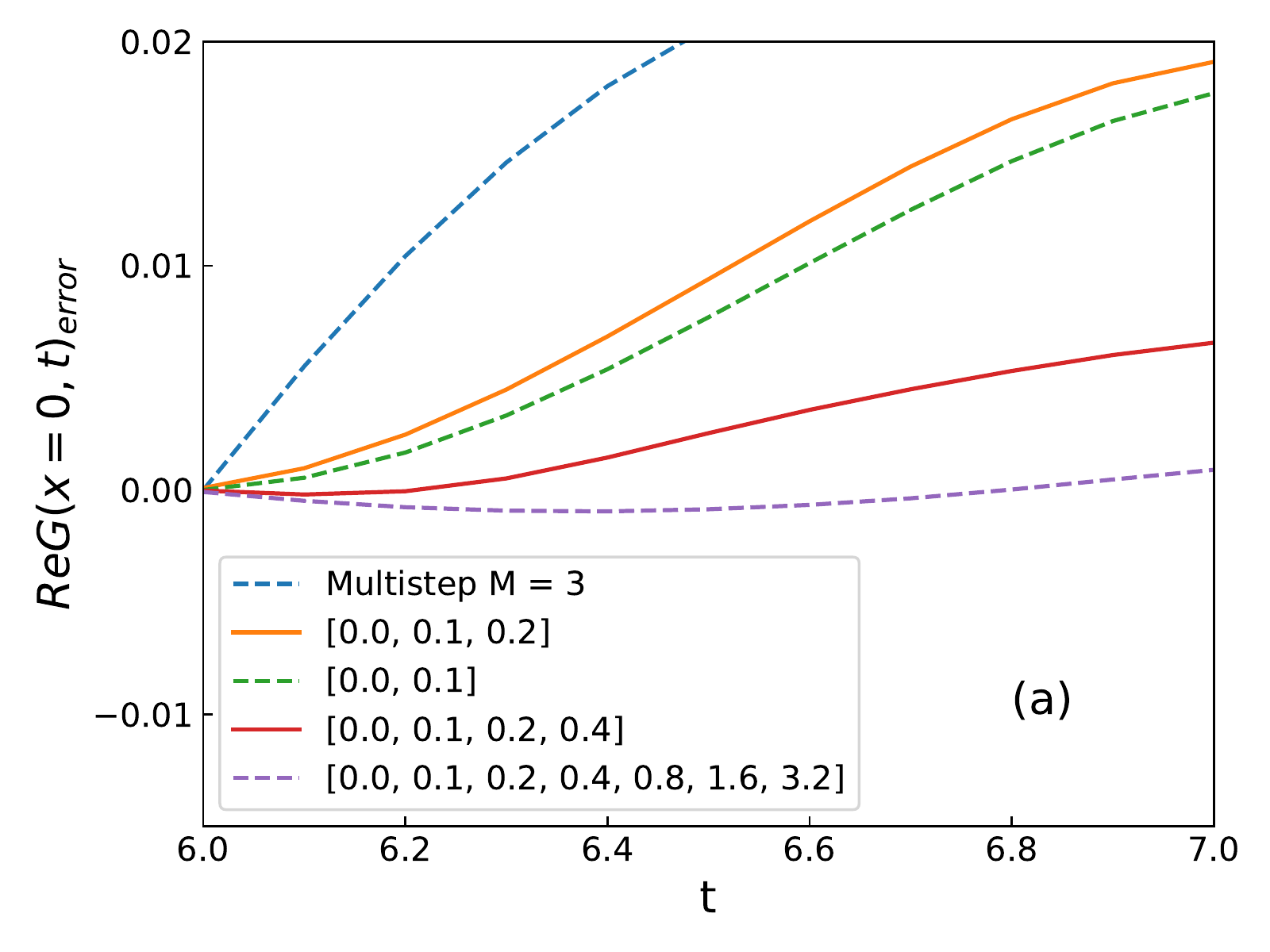}
\includegraphics[scale=0.37]{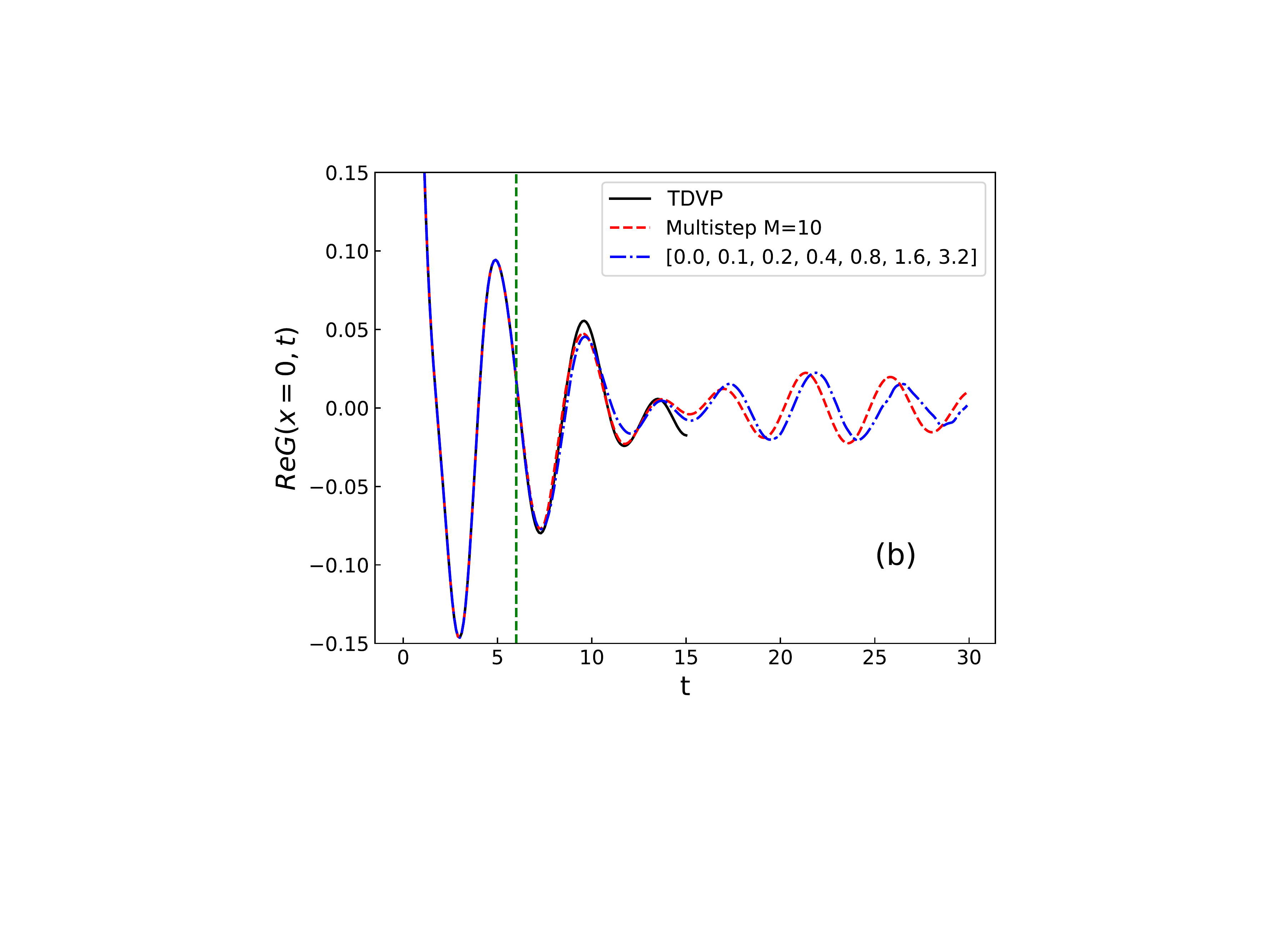}
\caption{Errors in the extrapolation of the on-site Green's function, relative to running the DMRG longer, for the MTB method, and compared with multi-step recursion. Part (a) shows the errors in the Green's function, while (b) shows the Green's function itself. The list in the legend shows the times for which the wavefunction is added into the basis, and the ordering of lines in the legend matches the height of the curves. The last (or only) projection was done at $t_1=6$ in all cases.  In (a), the multi-step recursion has spacing of 0.1 with 3 projections, while in (b),
we show results for 10 projections at a spacing of 0.5. }
\label{fig:expand_basis}
\end{figure}

Another variation on multistep recursion involves building into a single basis the wavefunctions from more than one time.  For example, we could make a basis out of the wavefunction and its translations at times $t=0$, $t=0.1$,
and $t=0.2$, making a basis and $O$ three times as large. We call this the multi-time basis method (MTB). This method is particularly good at removing the slope discontinuity.  The Green's functions at nearby times contain information about the short-time time evolution, and by combining them in the same basis, one or more derivative discontinuities can be eliminated, approximately.  Sample results for this method are shown in Fig. \ref{fig:expand_basis}. We have only briefly studied this method, and likely other variations are possible, which we leave for future work. 

\section{\label{sec:two} Time Dependent DMRG and linear prediction}

Starting from the Lehmann representation, a convenient formula for the spectral function, in 1D with a straightforward generalization to a ladder, is 
\begin{align} 
\begin{split}
A(k,\omega)
&= \sum_x e^{-ikx}
\frac{1}{2\pi}\Big{[}\int_{-\infty}^{\infty}e^{i(\omega-E_0) t} dt \langle 0|c_x(t)  c^{\dagger}_0|0\rangle
\\
&+ 
\int_{-\infty}^{\infty}e^{-i(\omega+E_0)t} dt \langle 0|c^{\dagger}_x(t) c_0|0\rangle\Big{]}
\end{split}
\label{spectral_eq}
\end{align} 
where $|0\rangle$ is the ground state.
The two pieces of the Green's function, $\langle0|c_x(t)  c^{\dagger}_0|0\rangle$ and $\langle 0|c^{\dagger}_x(t) c_0|0\rangle)$, need separate DMRG runs to compute. 
Taking the particle addition term as an example, let $ \left|\psi(t=0)\right\rangle \equiv c_0^\dagger\left|0\right\rangle$ and evolve this state with the operator $ e^{-i(H-E_0)t}$, where $E_0$ is the ground state energy. Then the desired Green's function is $G(x,t) = \left\langle0\right|c_x\left|\psi(t)\right\rangle$.  

We study the 2-leg ladder Hubbard model with Hamiltonian
\begin{equation} 
H=-t\sum_{\langle i,j \rangle,\sigma}c^\dagger_{i,\sigma} c_{j\sigma} + U\sum_l n_{i,\uparrow}n_{i,\downarrow} .
\end{equation}
where we sum over all nearest neighbor pairs of sites $\langle ij\rangle$ and $\sigma$ labels the spin. We set $t=1$ throughout, and all energy units are thus in terms of $t$, and time units in terms of $1/t$.
Our ground state DMRG calculation is quite standard, using a snake-like path through the ladder.\cite{Steven_1992} All the calculations reported here were performed using the ITensor library(http://itensor.org).  The ground state calculations kept 400-1000 states, enough to limit the truncation error per step to $10^{-7}$, with dozens of sweeps used to give good convergence. After adding or subtracting a particle at the center of the system, we evolved in time using the time depedent variatonal principle two-site algorithm (TDVP).  We also compared TDVP with the tDMRG algorithm\cite{Feiguin_2004},
which is equivalent to the time evolving block decimation algorithm\cite{TEBD_1}. For the tDMRG calculations, the ladder mandated the use of swap gates to allow all bond operators to be nearest neighbor in the MPS. The tDMRG calculations were identical to the approach described in \cite{White2015}. The efficiency and accuracy of these two methods is described in the Appendix. In general, we found the TDVP method to be more satisfactory, and our results primarily use that method. For the TDVP we kept enough states to attempt to achieve a truncation error of  $10^{-7}$,
but did not let the number of states go over 2000.  These parameters gave accurate results out to about $t=12-15$.

As time evolves, the wavepacket spreads out. We always stop the simulation at a time $t_{max}$ before the packet reaches the edges of the system. Thus any finite size effects are small.  We used a $64\times 2$ system, which was big enough for $t_{\rm max}=15$.

We compared our recursion methods with linear prediction\cite{Numerical_Recipes}.
Linear prediction extrapolates a discrete equally spaced time series ${y_i}$ as
\begin{equation} 
\tilde{y_i} = \sum_{j=1}^p a_j y_{i-j}
\end{equation} 
where $\tilde{y_i}$ is the predicted value.
The coefficients $a_j$ are determined by the known data points ${y_i}$ by requiring that their prediction for each point $y_i$, based on $y_{i-n}...y_{i-1}$, vary as little as possible from the actual values $y_i$, using a least-squares criterion. Linear prediction can be unstable, but there are standard procedure for correcting the coefficients of the recursion to enforce stability.

Linear prediction, in principle, works exactly on a small number of sine waves, or sine waves with exponential decay.  
Linear prediction methods are closely related to power spectral density estimation\cite{Numerical_Recipes}. 
The superposition of damped oscillating terms in linear prediction is determined by a distribution of poles in the complex plane, while power spectral density estimation is used to find these poles. 
We explored several methods for power spectral density estimation and found the Burg method\cite{SPECTRAL_ANALYSIS} is the most satisfactory. The Burg method minimizes the forward and backward prediction errors in the least squares sense, with advantages in that it gives a high resolution for short data records and always produces a stable model. 

In practice, linear prediction only has a limited amount of data to learn the signal, so it helps if the signal is fairly simple, say with one or a few dominant oscillations.
For this reason, it is best to Fourier transform $x \to k$ first, before applying the prediction, in obtaining a
spectral function $A(k,\omega)$. If one is interested in the total density of states $N(\omega)$, which is usually more complicated, it is best to perform linear prediction over all momenta, separately,  and then integrate over momenta. Linear prediction can completely fail if applied to $G(x,t)$ for fixed $x$ away from the origin, since the signal suddenly turns on when the wavefront reaches $x$. In contrast, our recursion methods work on the entire Green's in one
calculation, making use of all the data, and giving $N(\omega)$ and all desired $A(k,\omega)$.




\section{\label{sec:four} Multi-step recursion versus linear prediction}

\begin{figure}[h]
\includegraphics[width=1.0\linewidth]{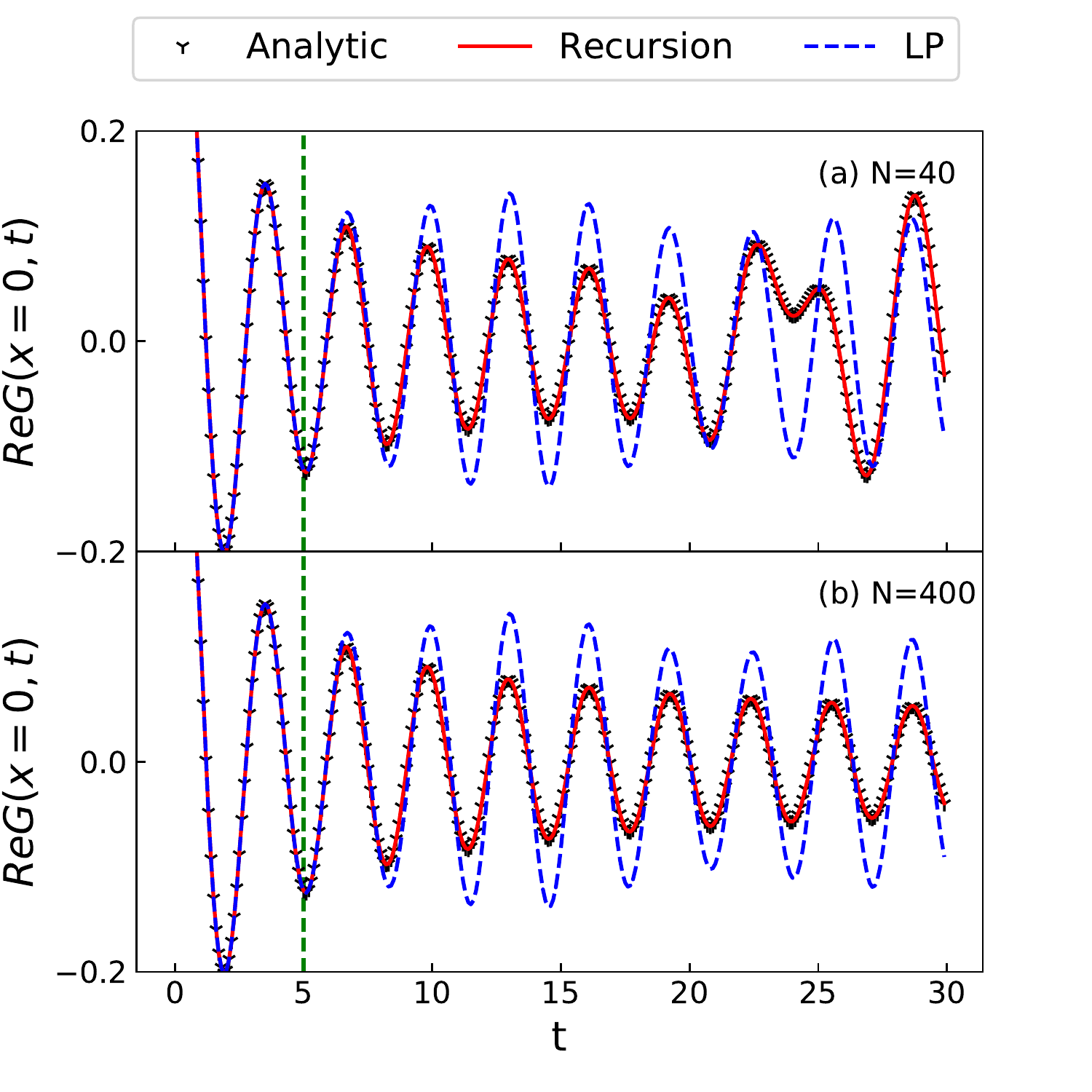}
\caption{Recursion results for the real part of the Green's function $G(x=0,t)$ for the 1D Hubbard model at U=0, N=40 at half-filling, (a), and U=0, N=400 at half-filling, (b). The recursion for the N=40 system is performed exactly without translational invariance. The black curve represents analytic data out to $t = 30$; the red curve is the analytic data for $t \leq 5$ and a recursion extrapolation for $t > 5$; the blue curve is the analytic data for $t \leq 5$ and a linear prediction extrapolation for $t > 5$.}
\label{fig:1D_U0}
\end{figure}

\begin{figure}[h]
\centering
\includegraphics[width=1.0\linewidth]{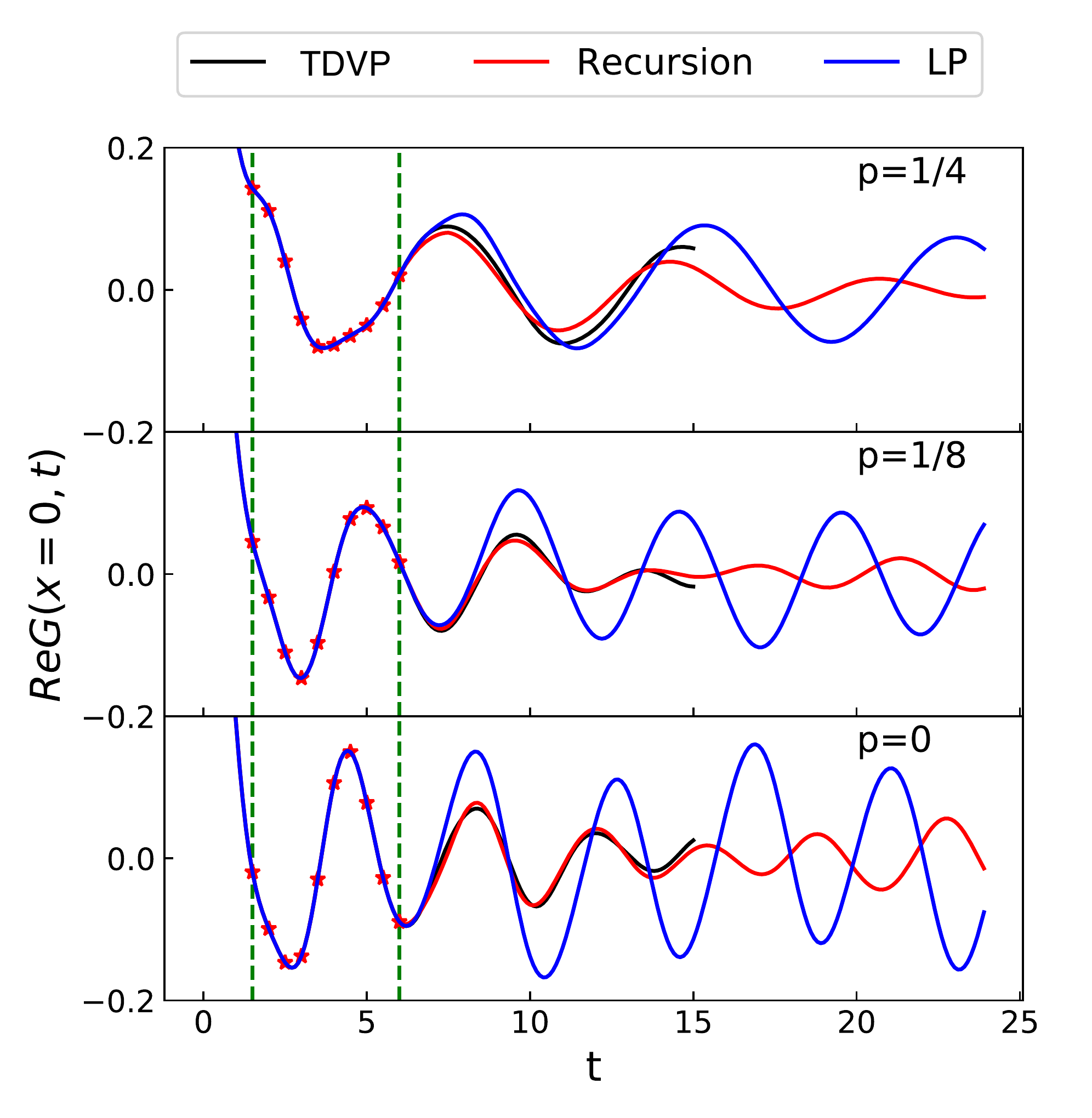}
\caption{The real part of the single particle Green's function $G(x=0,t)$ for the Hubbard model at U=8 on a $64\times2$ lattice at half-filling, $1/8$ hole-doping, and $1/4$ hole-doping, comparing multi-step recursion with linear prediction. In the recursion, the spacing is 0.5, and the number of projections is 10. The green vertical lines indicate the first and the last recursion times, while the red stars indicate all the times.}
\label{fig:ladder_half_U8_doping}
\end{figure}

\begin{figure}[h]
\centering
\includegraphics[width=1.0\linewidth]{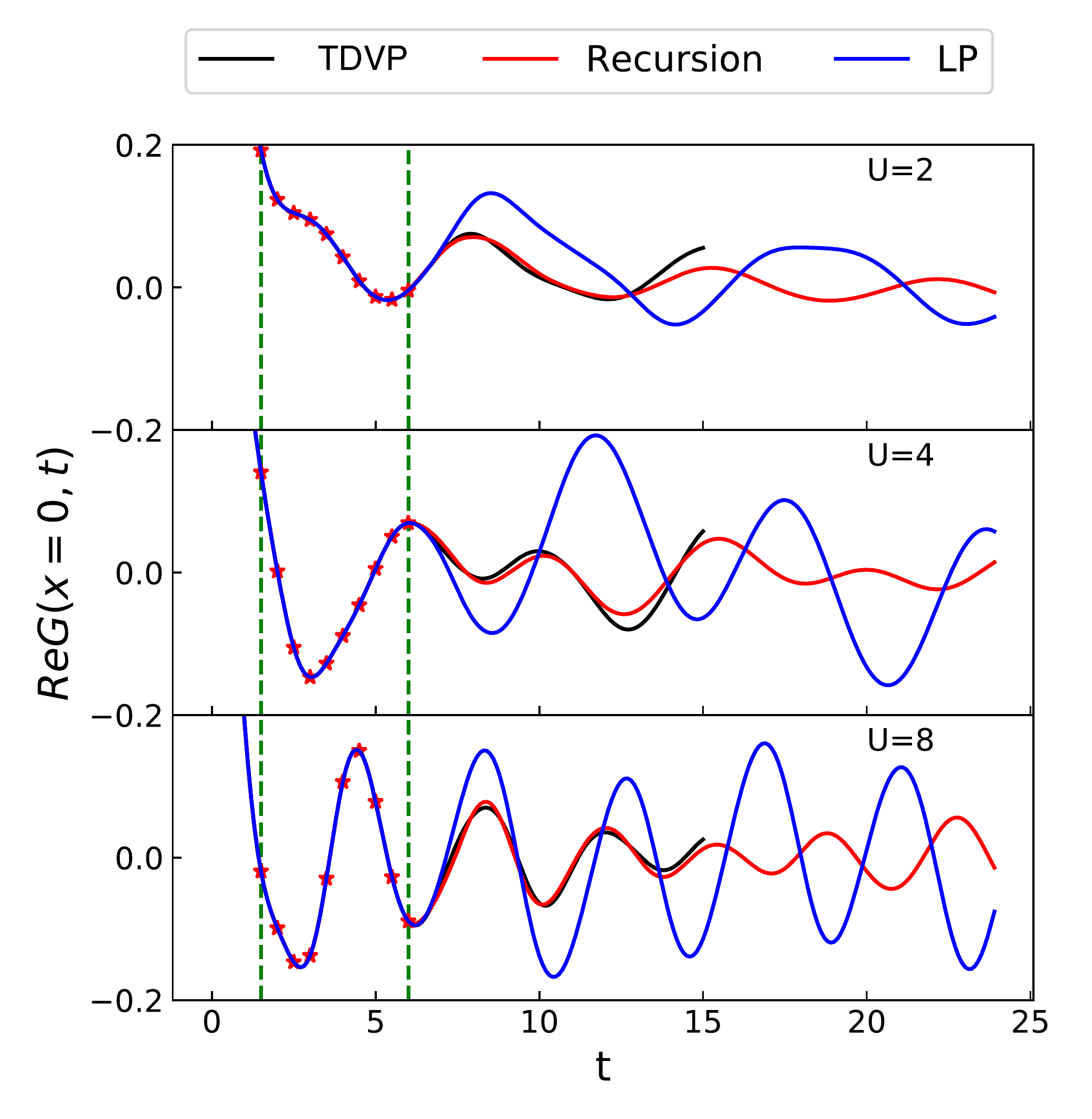}
\caption{The real part of the single particle Green's function $G(x=0,t)$ for the  Hubbard model at $U=2$, $U=4$, and $U=8$ on a $64\times2$ lattice at half-filling. The style is the same as in Figure~\ref{fig:ladder_half_U8_doping}.}
\label{fig:ladder_half_diff_U}
\end{figure}

In this section, we compare the recursion methods and linear prediction for extrapolating the Green's functions of the  1D  and 2-leg-ladder Hubbard model.  We first explored the recursion method on the 1D lattice at U=0 to verify that the recursion can yield the exact Green's function in the noninteracting case. Figure~\ref{fig:1D_U0}(a) shows the Green's function of 1D  model with N=40 at $U=0$ using one-step recursion with the computation of the full   $O_{ij}$ and $G_{ij}$ matrices, without using translational invariance.  In this case, the wavefront hits the edges of the system and bounces back, leading to irregular behavior near $t=25$, which the recursion can capture.  Figure~\ref{fig:1D_U0}(b) shows similar results on a longer system, where for the time of interest the wavefront has not hit the edges. In this case, we use our approximate translational invariance in the recursion.  The error in the extrapolation using recursion is not visible even with the translational invariance approximations. Comparing the two sizes, the initial signal is very similar, indicating that the primary finite size effects are due to the wavefront hitting the edges, and as long as one stops using the data before that time, finite size effects should be minimal. The extrapolation using linear prediction is reasonable but shows sizable errors.

Next, we compare Green's functions at finite $U$, where we perform an extrapolation using only an initial part of the DMRG data, and then compare with the  DMRG at later times. The amount of time we can go with DMRG is limited, but we can extrapolate longer, and check for instability.
Figure~\ref{fig:ladder_half_U8_doping} compares the extrapolation of Green's function $G(x=0,t)$ using the multistep recursion and linear prediction with the DMRG data up to t=6 at different hole-doping conditions at fixed U=8. We use 10 projections during the recursion procedure, the spacing between the two projections is 0.5. Figure~\ref{fig:ladder_half_diff_U} shows similar results at half-filling for different U. 
The difference between the DMRG and recursion is very small in Figure~\ref{fig:ladder_half_U8_doping} and  Figure~\ref{fig:ladder_half_diff_U}. In general, 
The recursion method very accurate at different doping and at different $U$ while linear prediction is less accurate.


The ladder is symmetric under reflection symmetry which interchanges the two legs; correspondingly, the $k_y$ component of the one-hole and one-particle states can be classified by their symmetry. We use the labels $k_y=0$ and $k_y=\pi$ to denote the even and odd symmetry modes, respectively. We get these two modes from a simulation where a particle is added (or removed) from one site in the center of the ladder. We first use the recursion method to extend the results to a larger time--typically about four times the length of the actual simulation. We then interchange legs in the Green's function to get an equivalent Green's function for when a particle is added to the other leg, and then recombine these two Green's functions to get the two modes.  Subsequently, we Fourier transform (FT) $x \to k_x$  to get  $G(k_x,k_y,t)$. We Fourier transform $t\to\omega$ using a Gaussian window $e^{-8*(t/t_{max})^2}$, where $t_{\rm max}$ is the maximum extrapolated time. This window leaves a negligible discontinuity at $t_{\rm max}$. The result is $A(k,\omega)$.

\begin{figure}[h]
\includegraphics[width=1.0\linewidth]{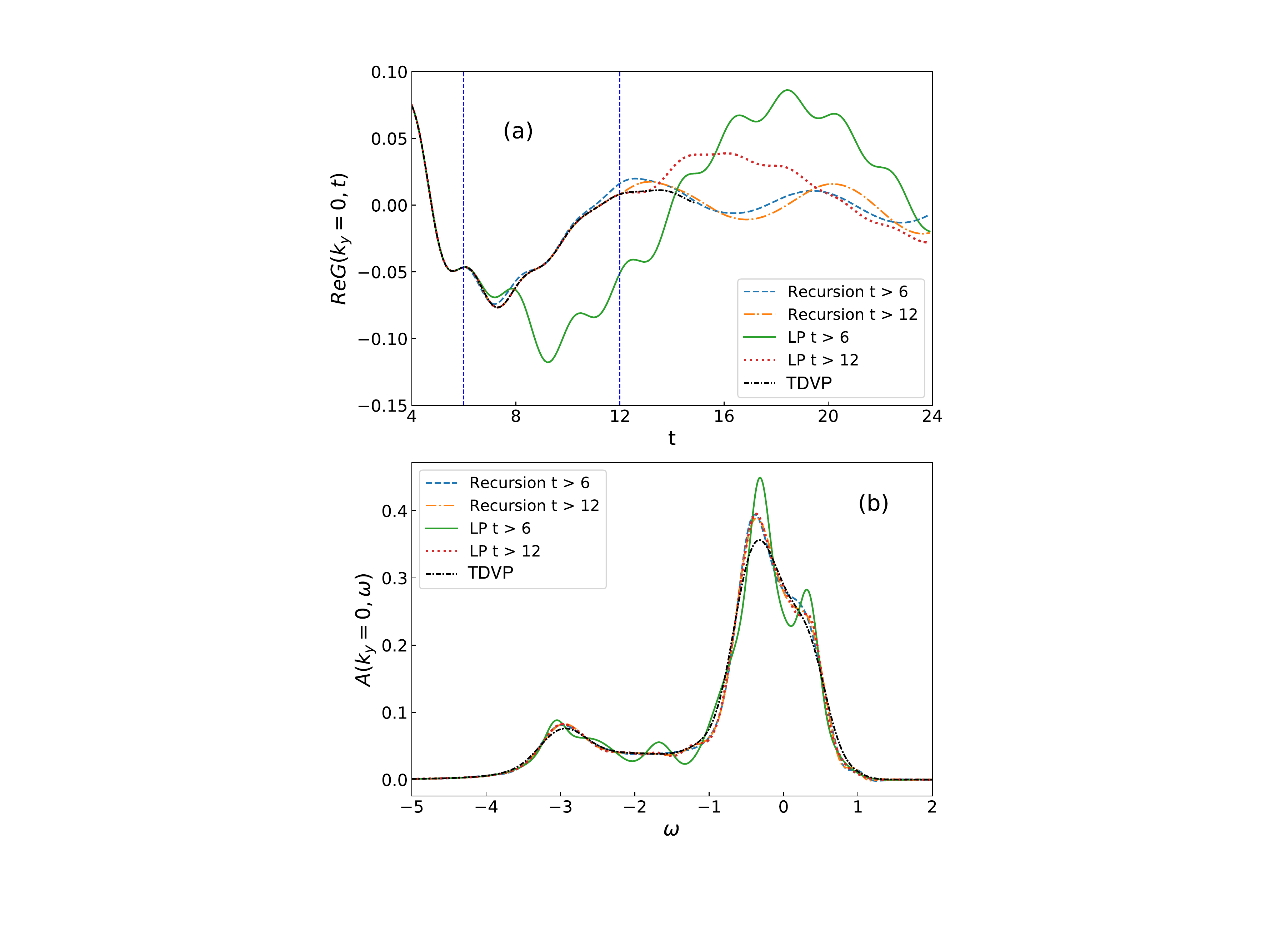}
\caption{(a) shows the real part of Green's function $G(k,t)$ at $k_x=0.3\pi$ for $64\times2$ ladder at U=8 and 1/4 hole-doping using multistep recursion and linear prediction for t>6 and t>12. The black line is the DMRG data. The $t_{max}$ is 24. (b) shows the spectral function $A(k,\omega)$ by applying FT to the data from the (a).}
\label{fig:LP_ladder_4th_U8_Greenk_compare}
\end{figure}

At small $U$ and when the spectrum is dominated by sharp peaks, both linear prediction and recursion have an easier time extrapolating the data.
Away from this regime, both methods have more difficulty, especially linear prediction, if the maximum time is small.  Figure~\ref{fig:LP_ladder_4th_U8_Greenk_compare}(a) shows an example of this.  If simulation data were only available out to $t=6$, the linear prediction extrapolation at this point, $k_x=0.3\pi,k_y=0$ for $U=8$ at $1/4$ doping, immediately deviates strongly from the correct behavior.  Note that in this case, for linear prediction, we FT $x\to k_x$ before linear prediction, to make the signal as simple as possible. If data is available out to $t=12$, the linear prediction is better but still inadequate. The recursion, in contrast, does rather well for either range of times.  Figure~\ref{fig:LP_ladder_4th_U8_Greenk_compare}(b) shows the resulting spectral functions.  Since the extrapolations have the time window applied to them, the results are not as poor as one might expect for the $t>6$ linear prediction, but there are still significant oscillations and false peaks.  The multi-step recursion results are quite good and consistent. The difference between these two methods and the superiority of multistep recursion is frequently observed for most  $k$ values. 

\begin{figure}[h]
\centering
\includegraphics[width=1.0\linewidth]{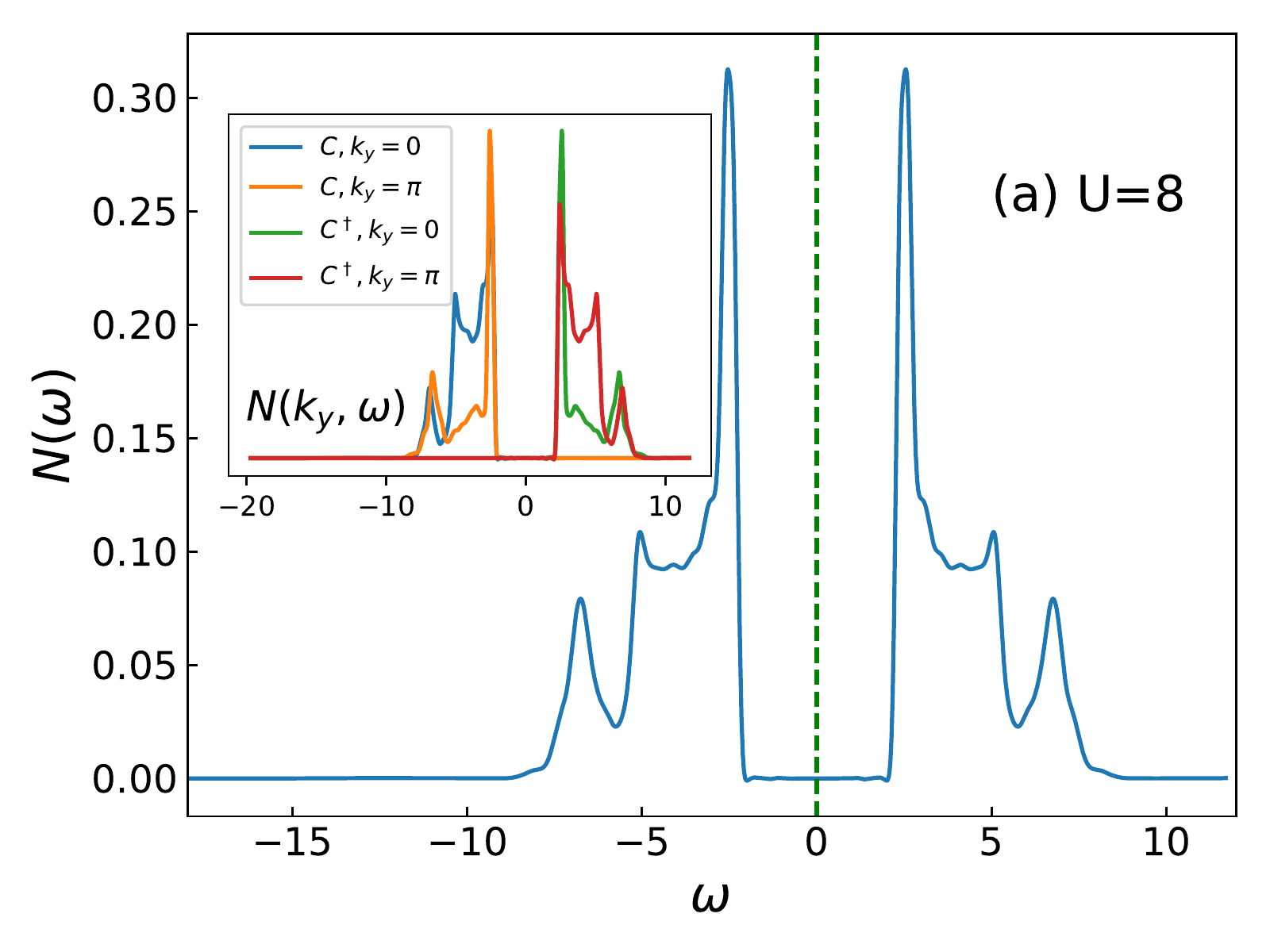}
\includegraphics[width=1.0\linewidth]{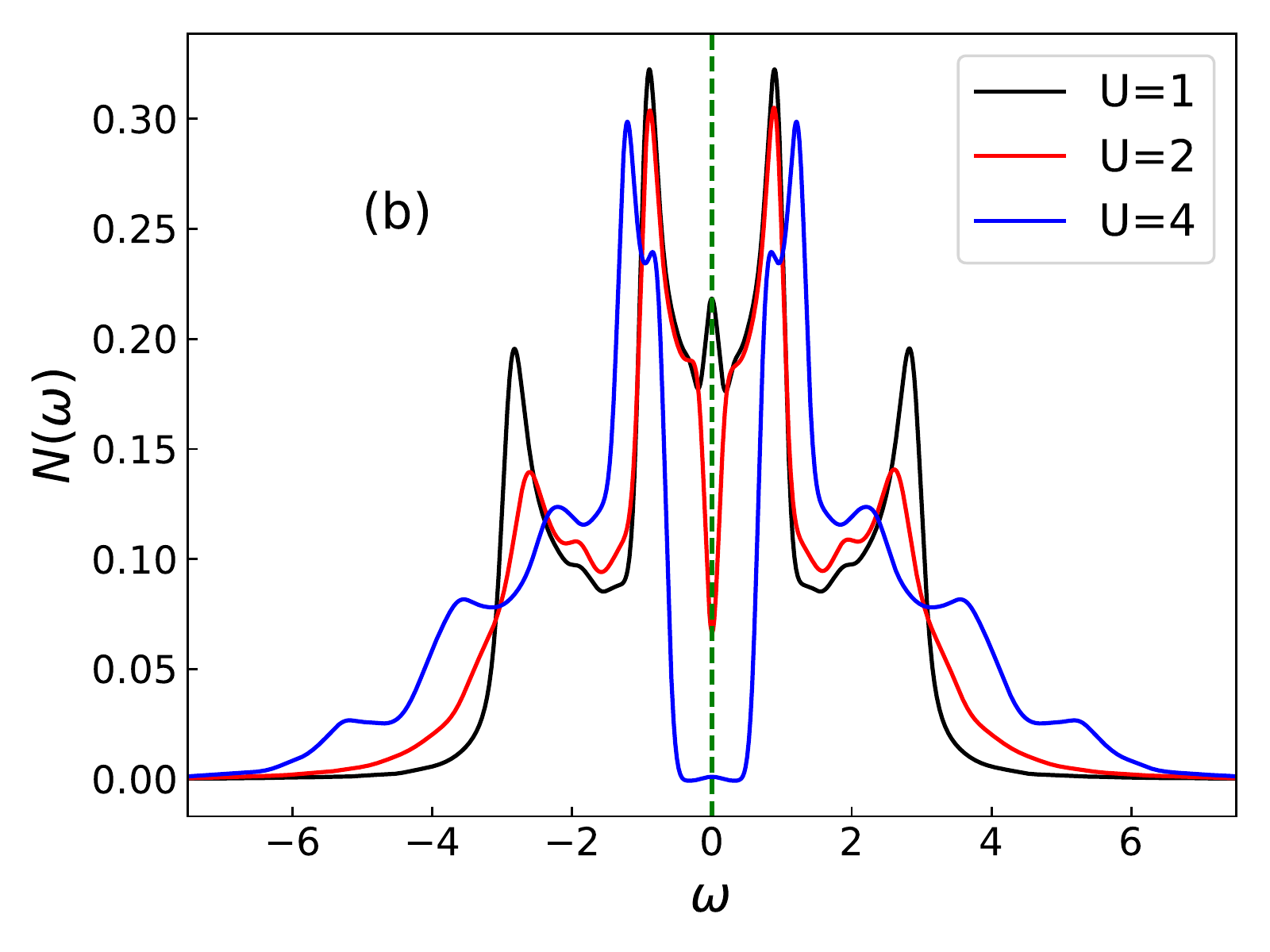}
\caption{Results for the density of states $N(\omega)$ for the $64\times2$ system at U=8 and half-filling with multistep recursion, utilizing the data from $t=10$ and extrapolate out to $t_{max}$ = 40. The data out to $t_{max}$ was multiplied by a simple Gaussian window with factor of 8. The inset shows the density of states of particle and holon of each $k_y$.}
\label{fig:ladder_half_density}
\end{figure}

\section{\label{sec:five} Spectral functions of the two leg Hubbard ladder}
\begin{figure*}
\centering
\includegraphics[width=1.0\linewidth]{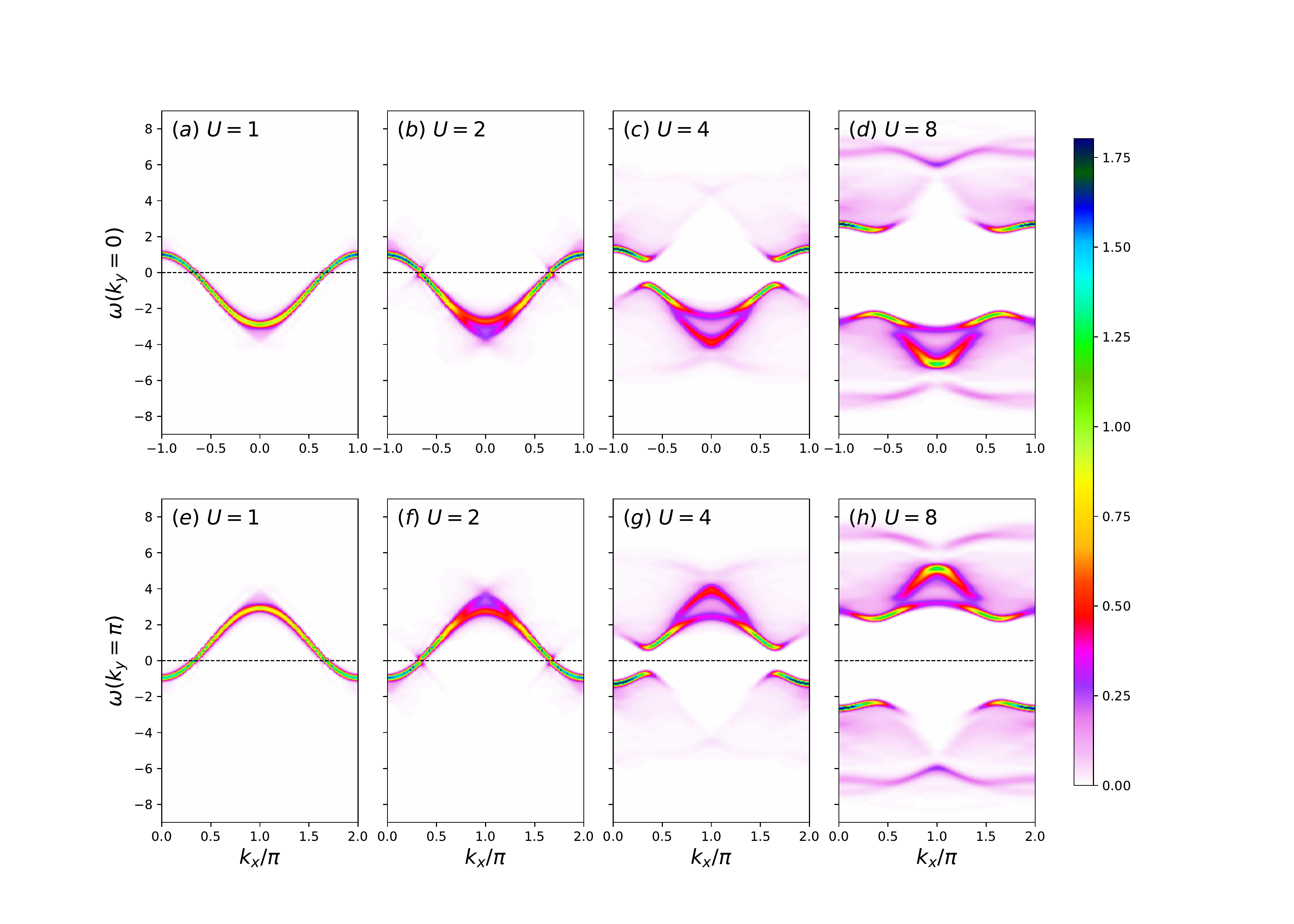}
\caption{The spectral function $A(k,\omega)$ at $k_y=0, k_y=\pi$ for Hubbard model on $64\times2$ lattice for various $U$ at half-filling. Note that in order to display the particle-hole symmetry, the upper panels for $k_y=0$ show $-\pi < k_x < \pi$ while the lower panels for $k_y=\pi$ show $0 < k_x < 2\pi$.}
\label{fig:ladder_half_spectral}
\end{figure*}

\begin{figure*}
\centering
\includegraphics[width=1.0\linewidth]{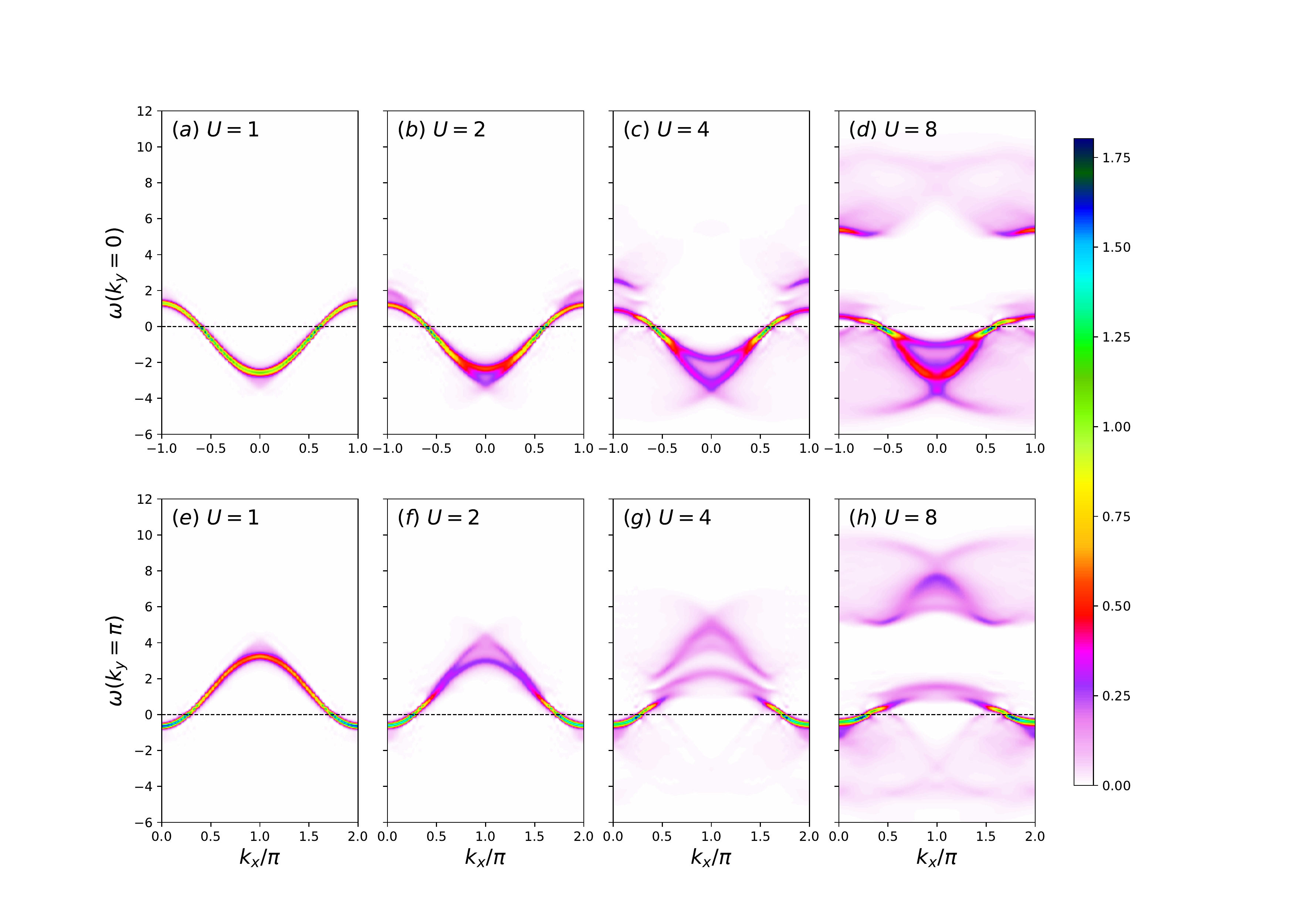}
\caption{The spectral function $A(k,\omega)$ at $k_y=0, k_y=\pi$ for Hubbard model on $64\times2$ lattice at U=8 at p=1/8. The style is same as Fig(\ref{fig:ladder_half_spectral}).  Here the dashed line at $\omega=0$ indicates the Fermi level. }
\label{fig:ladder_8th_spectral}
\end{figure*}

\begin{figure*}
\centering
\includegraphics[width=1.0\linewidth]{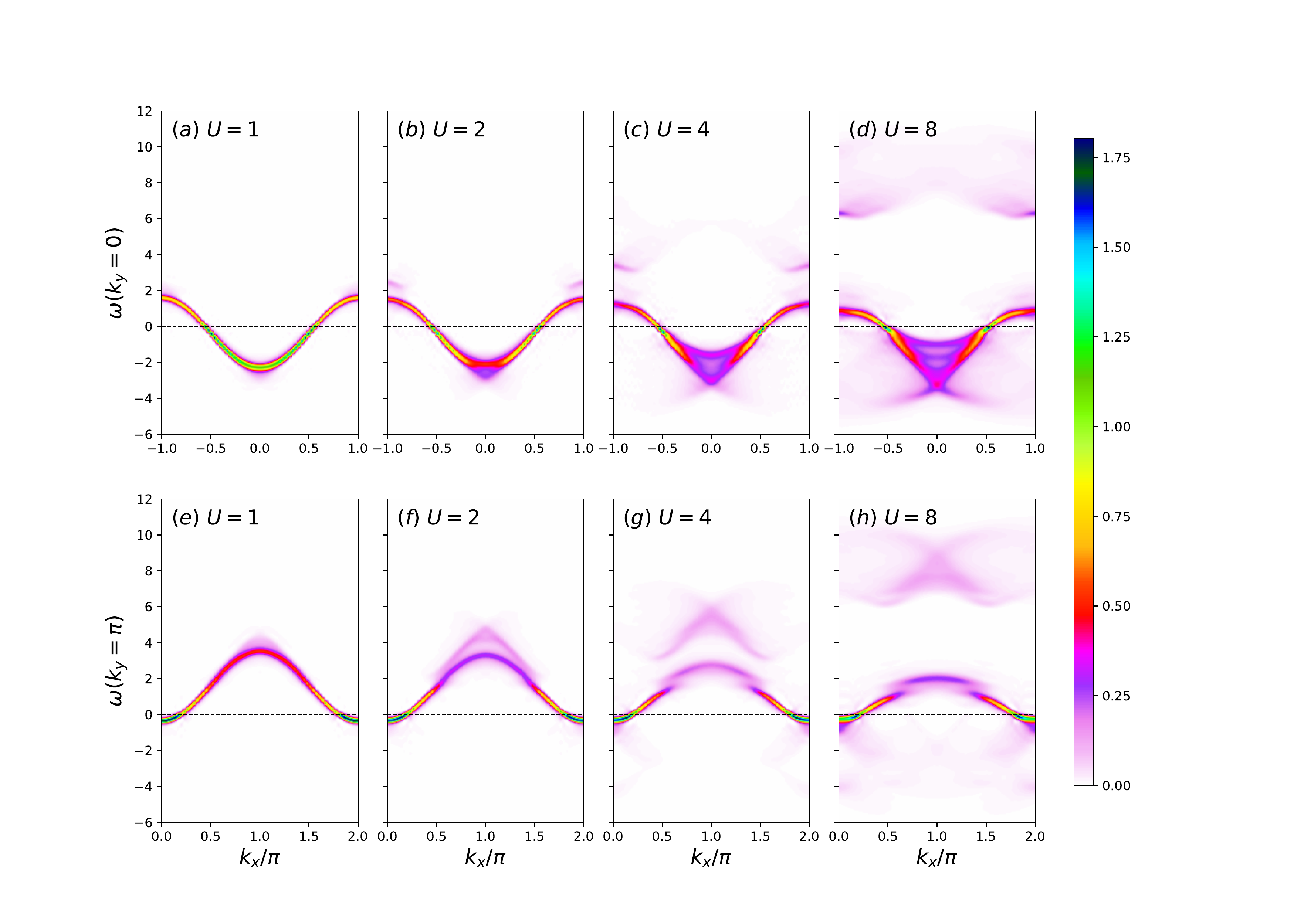}
\caption{The spectral function $A(k,\omega)$ at $k_y=0, k_y=\pi$ for Hubbard model on $64\times2$ lattice at U=8 at 1/4. Note that in order to display the particle-hole symmetry, the upper panels for $k_y=0$ show $-\pi < k_x < \pi$ while the lower panels for $k_y=\pi$ show $0 < k_x < 2\pi$.}
\label{fig:ladder_4th_spectral}
\end{figure*}



In this section we combine TDVP and multi-step recursion to present high quality spectral functions for the Hubbard ladder. We attempted to achieve a truncation error of $10^{-7}$ subject to a maximum number of states of $m=2000$. Although we simulated out to a time of 15, we found that there appeared to be a small loss of accuracy for the longest times which worsened the extrapolation, so we reduced the range of data used to 12. We then found that we could extrapolate out to about t=40 and the resulting spectra did not show significant unphysical oscillations. We show results using multi-step recursion with a time spacing of $0.5$ and $M=10$ recursions, but the multi-time basis results would be nearly identical. For each system, separate DMRG runs for different particle numbers   to determine the chemical potential, which was then used to set the Fermi energy to correspond to $\omega=0$. We found that there were some issues with instability, particularly at $U=4$, so we implemented the eigenvalue adjustment method described above, eliminating all instabilities. 

We note that even with recursion we are not able to resolve the smallest gaps.  In particular, the ladder at half-filling has a spin gap near 0.11-0.13 at $U=4-8$,
\cite{Noack1996} below our resolution, and any related single particle gaps of similar size also could not be resolved. It is not our intention to focus on the interesting nature of gaps of this size; instead we focus on the broad features at larger energy scales. Much work has been done on understanding the low energy properties of the ladder, for example\cite{Gannot2020,Orgad2001}.

We first consider the total density of states at half-filling, obtained from Fourier transforming the on-site Green's function (i.e. not integrating $A(k)$ over $k$), which was obtained separately from runs where a particle was added or destroyed. 
Figure~\ref{fig:ladder_half_density} shows our results. The particle-hole symmetry at half-filling is evident. A gap of about 4 is present at $U=8$, but the distance between the centers of the upper and lower bands is closer to $U$. Upon decreasing $U$, the gap shrinks, as shown in (b). At $U=2$, a significant dip is present, presumably reflecting a small but full gap, broadened by our finite resolution.  No sign of a gap is present at $U=1$; instead, a small peak at $\omega=0$ is present, but likely there is a very small gap that we cannot resolve. The upper and lower band have several peaks. We can get an initial idea of the origin of these peaks by decomposing the total density of states into particle and hole parts and also into the two transverse momenta, $k_y=0$ and $k_y=\pi$, which corresponds to even and odd modes on a rung. The inset to (a) shows this decomposition of $N(\omega)$ into these four parts.  We see that the particle-hole symmetry involves $\omega \to -\omega$ and $k_y \to k_y+\pi$. 
In terms of the spectral functions, the particle-hole symmetry is more precisely given by $A(k_x,\pi,\omega) = A(\pi-k_x,0,-\omega)$. Note that the two innermost and outermost peaks have contributions from both $k_y$ modes, but the smaller middle peaks only has contributions from one mode on each side.  Below, we see that these middle peaks are nevertheless quite sharp in the $A(k,\omega)$.

The next set of figures (Figs. \ref{fig:ladder_half_spectral}, \ref{fig:ladder_8th_spectral},\ref{fig:ladder_4th_spectral})  shows the spectral functions for several values of $U$ and several different doping.  The main features have mostly been seen before in other (mostly lower resolution) studies\cite{Feiguin2019,Martins1999}. Note that to more clearly show the particle-hole symmetry, we have used a different $k_x$ range in plotting the two different $k_y$ modes, but no alterations were made to the data itself. At $U=1$, the results are simple with very little modification to a single particle picture. One sees clearly the single particle bands, with little diffuse components. At $U=2$, at half-filling, one sees some broadening of the bands at the regions farthest from the Fermi surface, and somewhat more diffuse spectral weight.  A small feature is noticeable at the Fermi surface, corresponding to a gap at the limit of resolution for our calculations. At $U=4$, at half-filling, substantially more modifications to the simple band picture appear. The gap is fully evident.  Far from $\omega=0$, at the center of each $U=4$ panel, the band has also split into two lines, both broadened. Substantially more diffuse spectral weight.  Finally, at $U=8$ at half-filling, the spectral function bears little resemblance to noninteracting bands. Very substantial diffuse weight is present. There are several more line-like features, but most such lines are broadened. The lines are sharper at the edges of the gap, but only for some ranges of $k$. However, one feature has become much sharper than that at $U=4$, the peak at $k=(0,0)$ and its symmetry partner $k=(\pi,\pi)$.  This sharp peak gives rise to the small middle peaks on each side of $N(\omega)$ in Fig. \ref{fig:ladder_half_density}. 

At $1/8$ doping, many features of the spectral function are similar to half-filling, such as the general dependence on $U$.  The diffuse spectral weight rises sharply with $U$, but it is slightly reduced in overall magnitude compare to half-filling. The upper band is still visible but notably reduced in magnitude and also broader at $U=4$ and $U=8$.  Where the bands cross the Fermi surface, sharp peaks are present for all $U$, and for larger $U$, this is the only place where peaks are sharp. The sharp peak at $k=(0,0)$ seen at half-filling is now an undistinguished part of a broadened band.  At $1/4$ doping, these differences from half-filling increase.  There is still substantial diffuse scattering at $U=4$ and $U=8$, but the upper band is quite weak. Some features near $k=(0,0)$, below the Fermi surface, continue to have substantial structure.  All Fermi surface band crossings remain sharp.  The Fermi surface still intersects the $k_y=\pi$ band at $1/4$ doping. 

\begin{figure*}[h]
\centering
\includegraphics[width=0.47\linewidth]{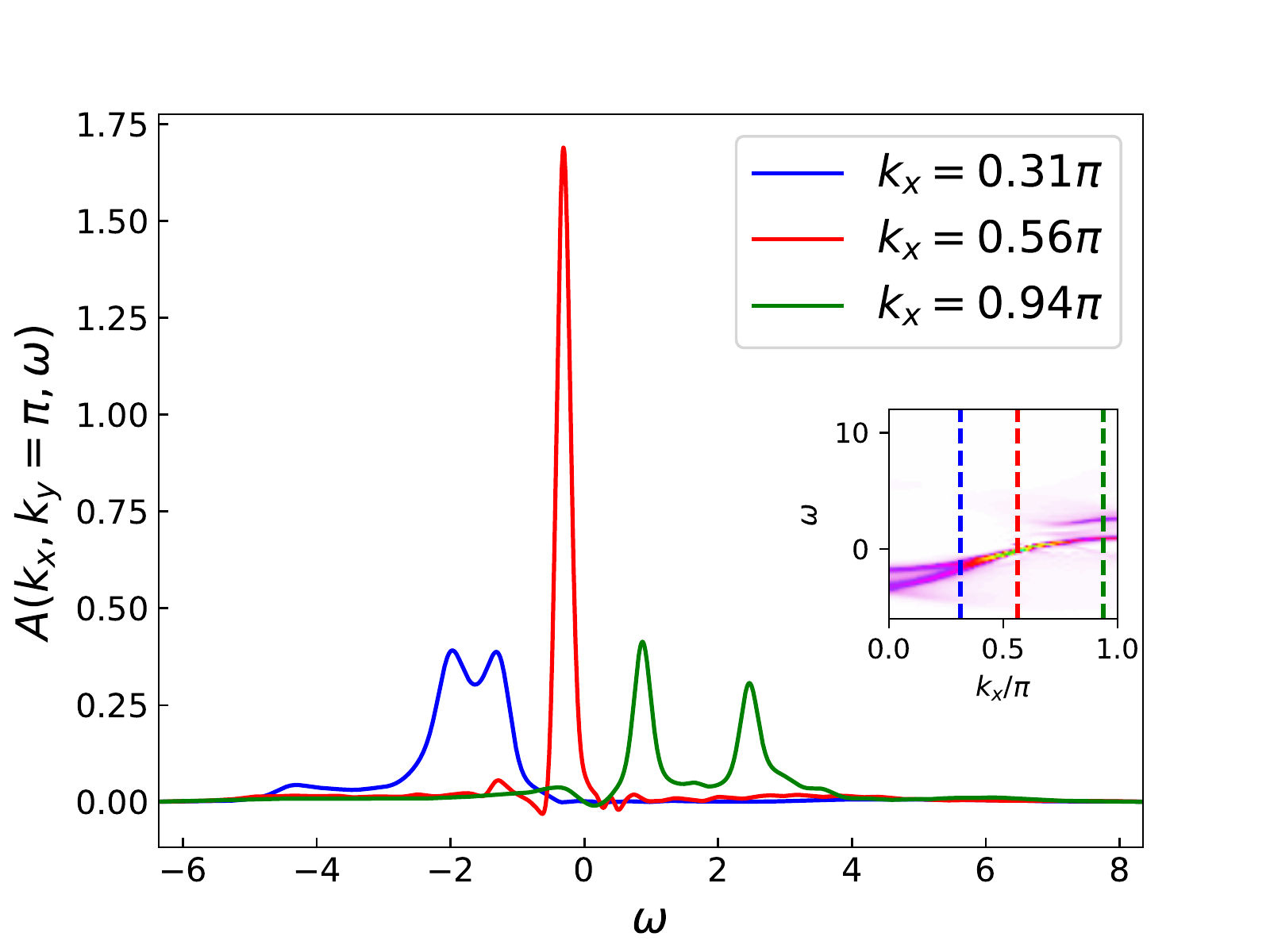}
\includegraphics[width=0.47\linewidth]{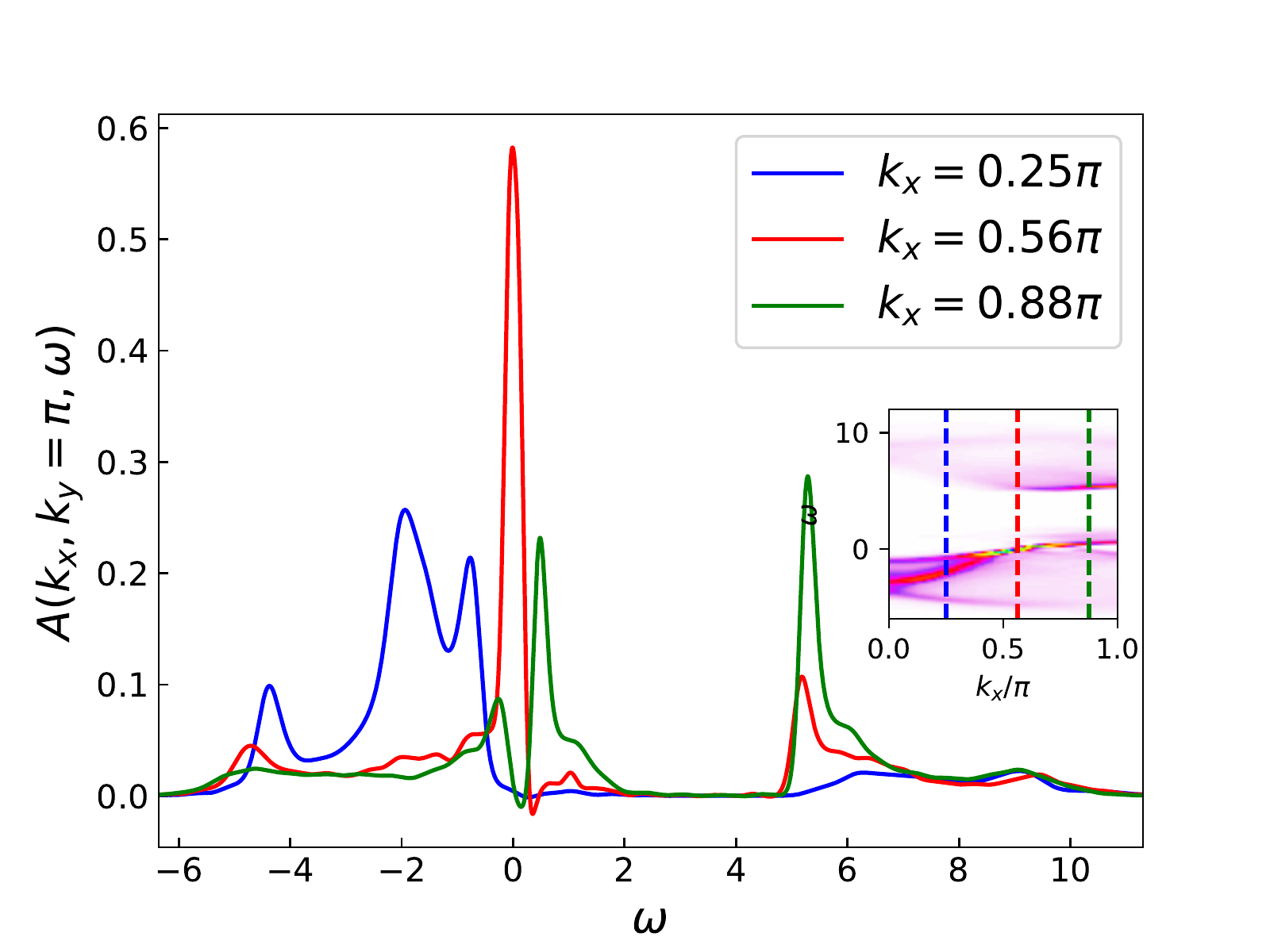}
\caption{(a) $A(k,\omega)$ at U = 4 with 1/8 hole-doping at $k_y=0$. The inset shows where the slice is. (b)$A(k,\omega)$ at U = 8 with 1/8 hole-doping at $k_y=0$.}
\label{fig:8th_doping_U4_U8_ky0}
\end{figure*}

\begin{figure*}[h]
\centering
\includegraphics[width=0.47\linewidth]{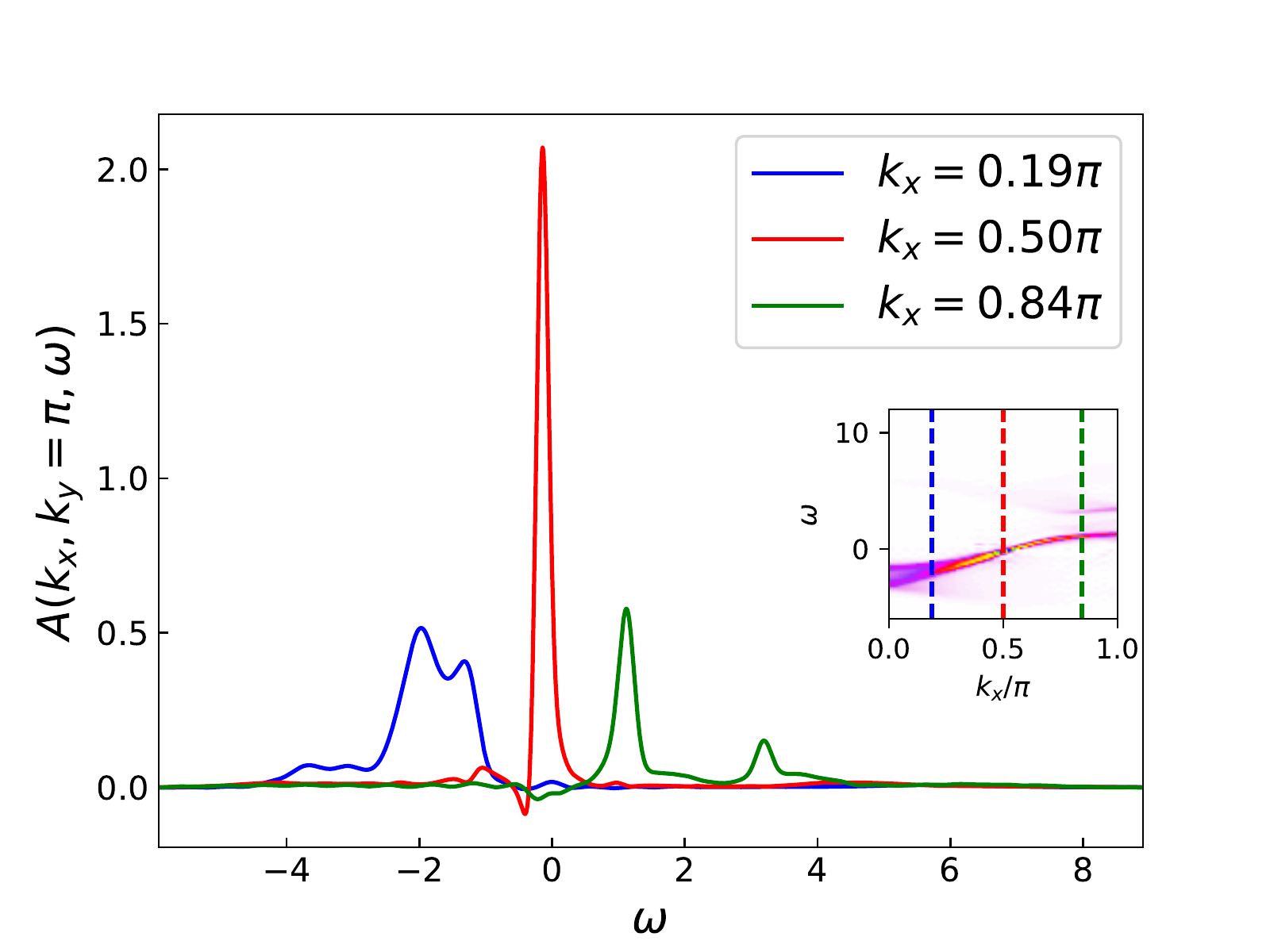}
\includegraphics[width=0.47\linewidth]{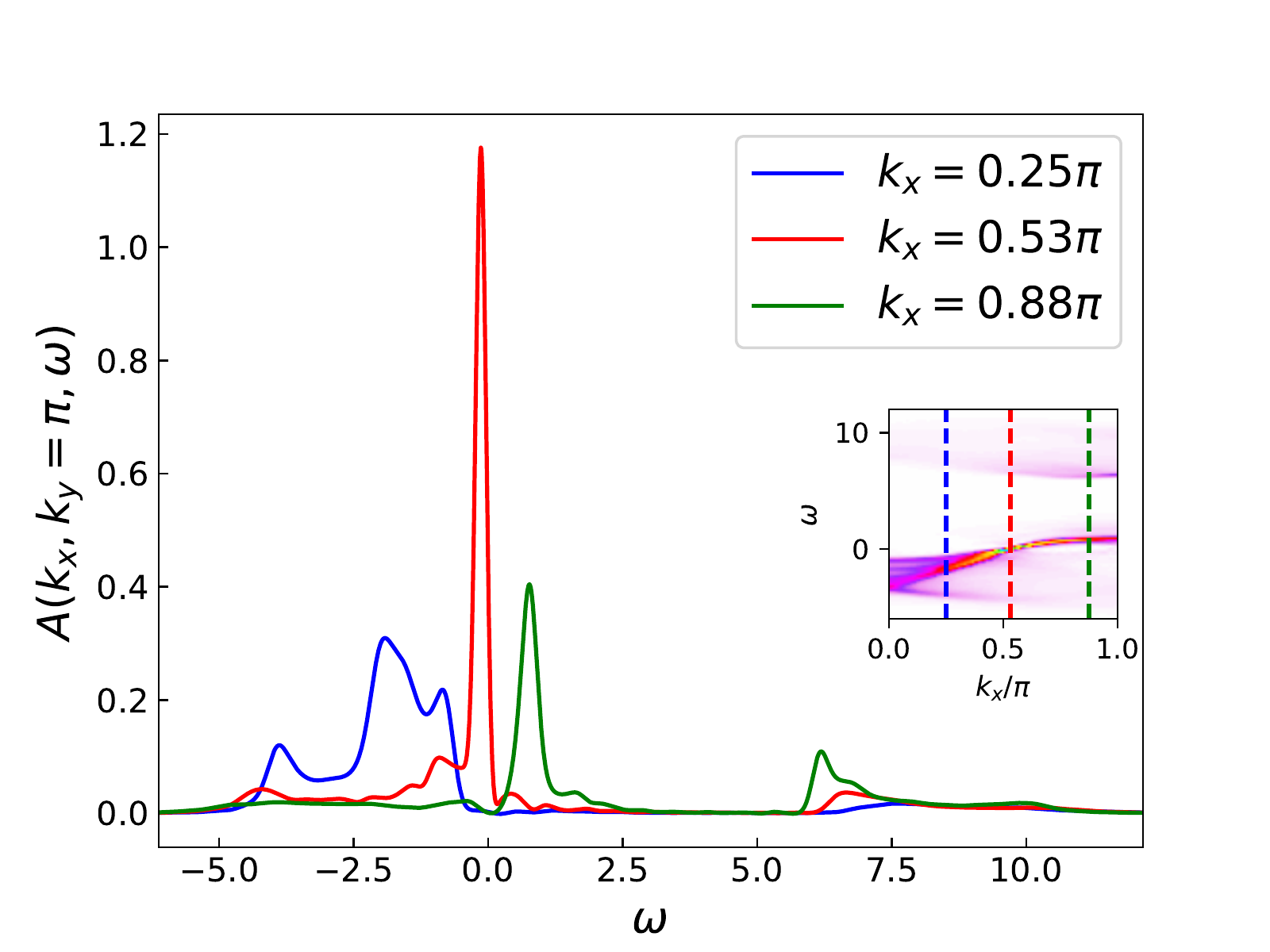}
\caption{(a) $A(k,\omega)$ at U = 4 with 1/4 hole-doping at $k_y=0$. The inset shows where the slice is. (b) $A(k,\omega)$ at U = 8 with 1/4 hole-doping at $k_y=0$.}
\label{fig:4th_doping_U4_U8_ky0}
\end{figure*}

\begin{figure*}[h]
\includegraphics[width=0.47\linewidth]{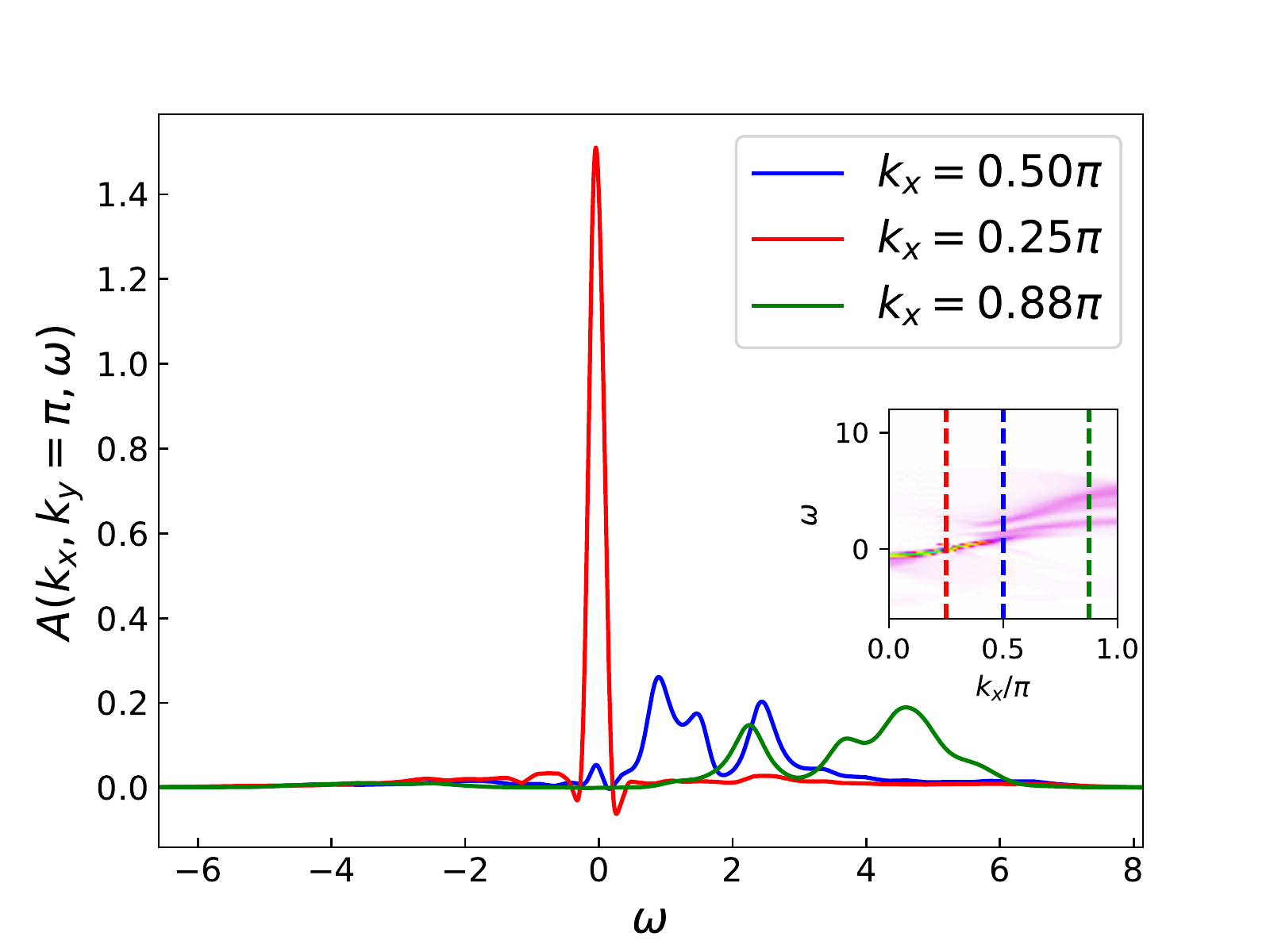}
\includegraphics[width=0.47\linewidth]{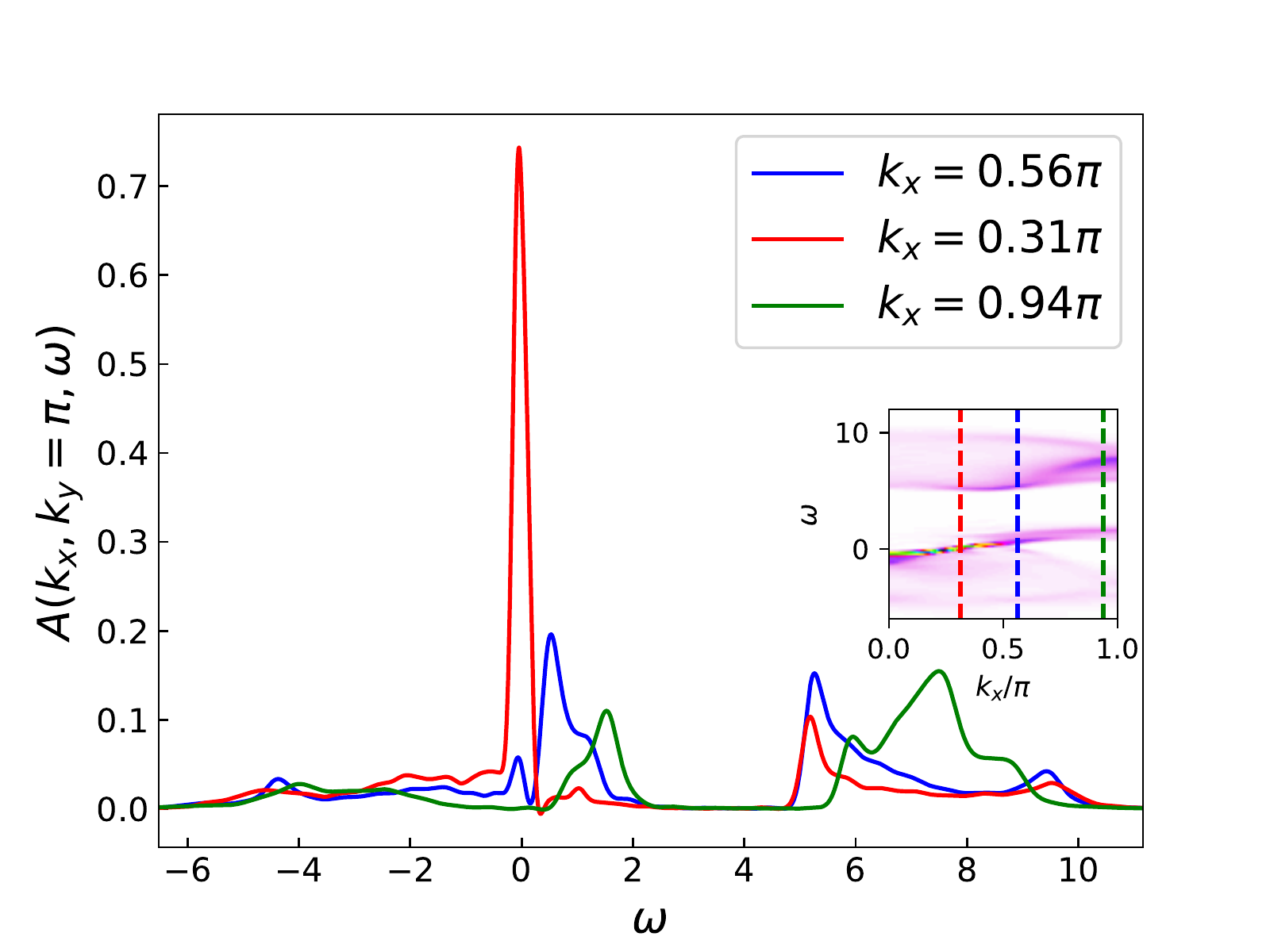}
\caption{(a) $A(k,\omega)$ at U = 4 with 1/8 hole-doping at $k_y=\pi$. The inset shows where the slice is. (b) $A(k,\omega)$ at U = 8 with 1/8 hole-doping at $k_y=\pi$.}
\label{fig:8th_doping_U4_U8_ky1}
\end{figure*}

\begin{figure*}[h]
\includegraphics[width=0.47\linewidth]{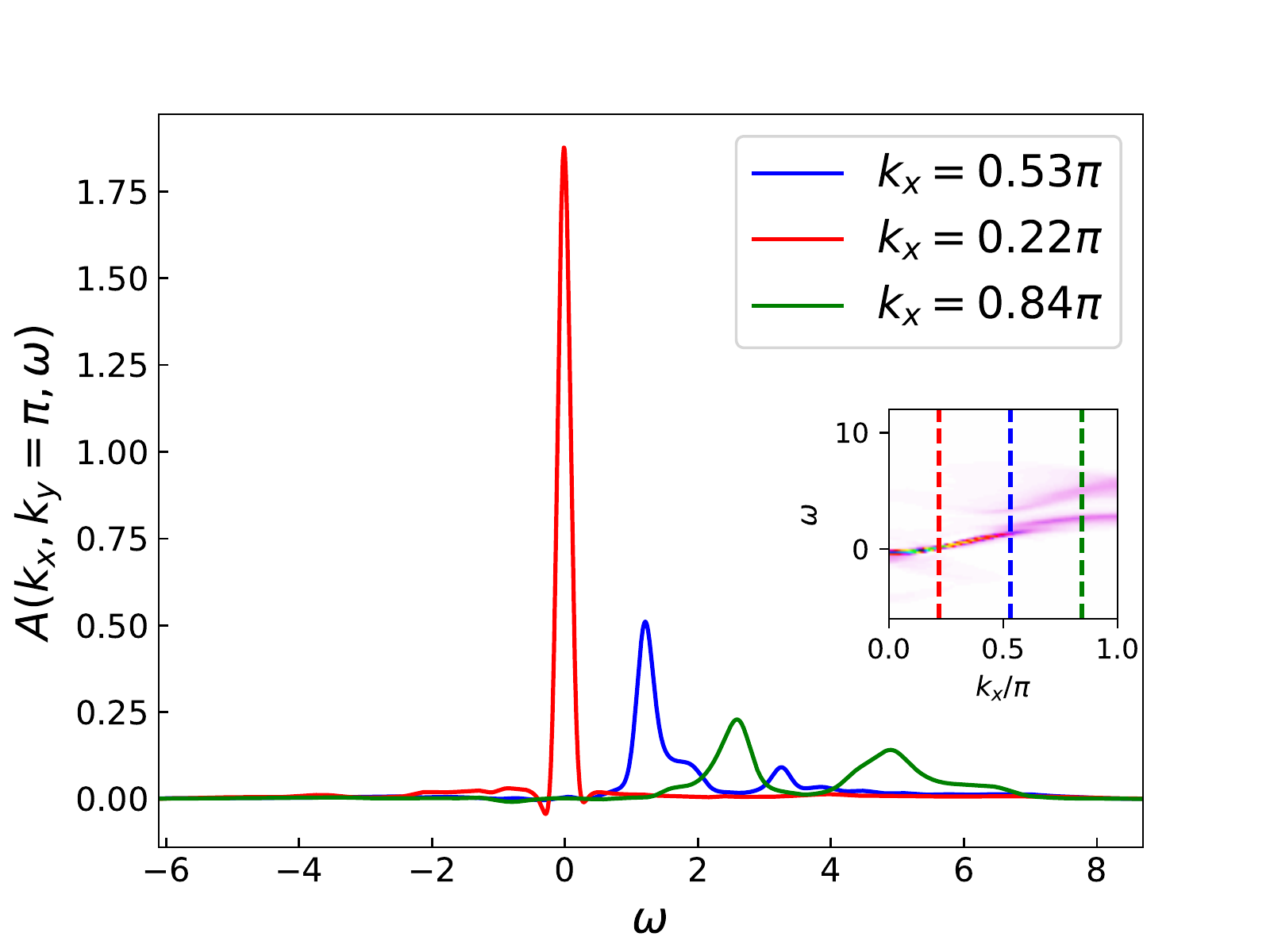}
\includegraphics[width=0.47\linewidth]{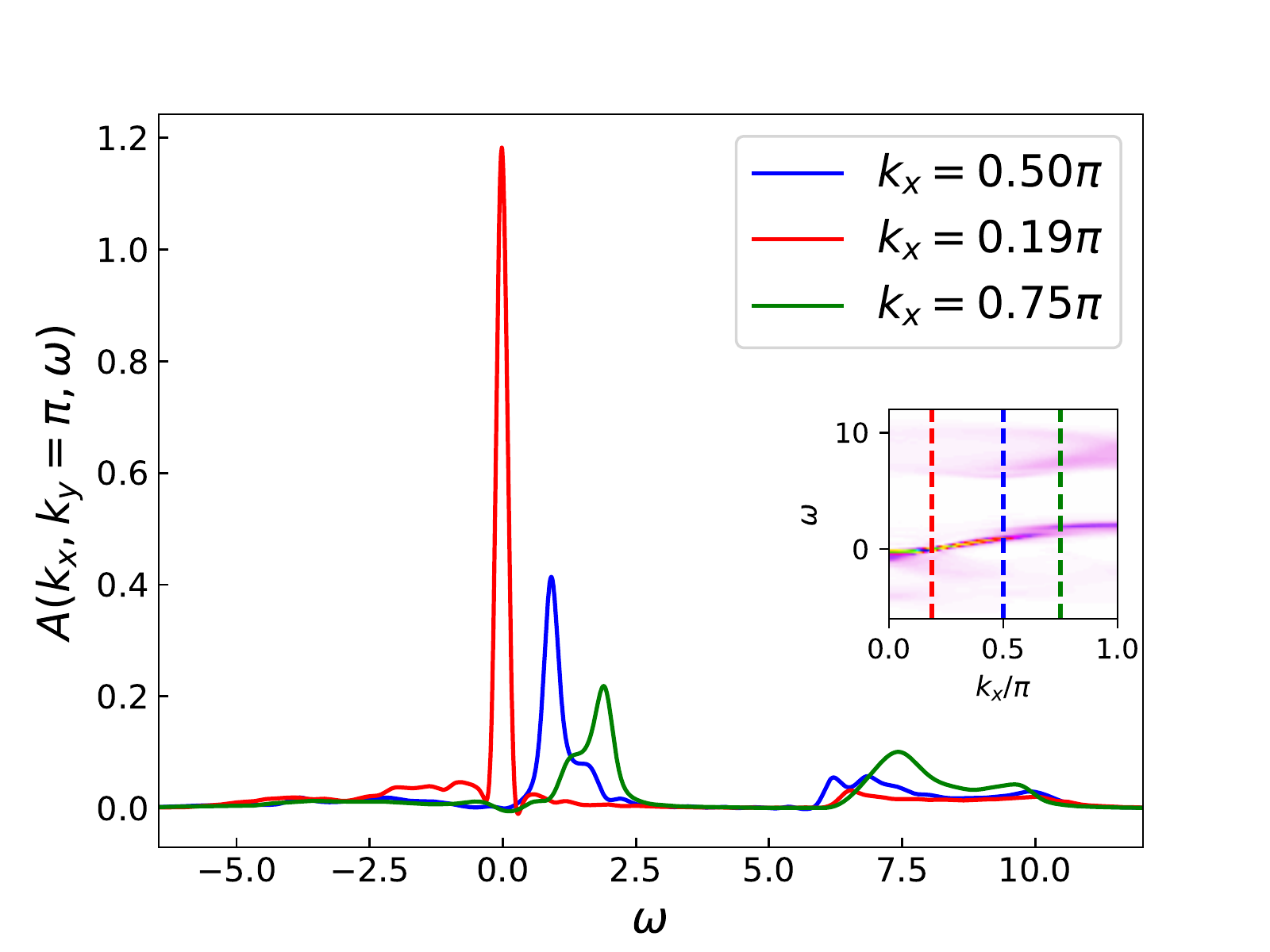}
\caption{(a) $A(k,\omega)$ at U = 4 with 1/4 hole-doping at $k_y=\pi$. The inset shows where the slice is. (b) $A(k,\omega)$ at U = 8 with 1/4 hole-doping at $k_y=\pi$.}
\label{fig:4th_doping_U4_U8_ky1}
\end{figure*}

To show the features more quantitatively, in Figs. \ref{fig:8th_doping_U4_U8_ky0} to \ref{fig:4th_doping_U4_U8_ky1} we show some ``slices'' of the spectral function for particular $k$, for doping $1/8$ and $1/4$, and $U=4, 8$.  In each case, we try to pick out the Fermi points and compare them with $k$-points above and below that. We see the sharpness of each Fermi surface peak, whose width is determined by the resolution of our calculations. The spectra at other $k$ values should have similar resolution, so the broadened multi-peak structure seen in the plots is generally clearly resolved.

\section{CONCLUSIONS}
In this article, we have introduced a set of recursion methods to extrapolate dynamical correlations functions coming from matrix product state solutions of the time-dependent Schrodinger equation, to allow improved calculation of spectral functions. We have compared these methods to linear prediction, which is currently the standard choice. We find that our multistep recursion method is generally more reliable and allows longer accurate extrapolations than linear prediction. We also describe a multi-time basis recursion method that performs similarly to multistep recursion, although with better accuracy at short times. For noninteracting Fermion systems, these methods are exact, but they perform well even in the strong coupling limit of the Hubbard model. 

We showcase the new approach with high-resolution calculations of the spectral functions of the two-leg Hubbard ladder, at half-filling and dopings of $p=1/8$ and $p=1/4$, for a range of $U$.  Our results provide substantial detail, such as the presence of weak diffuse spectral weight, and broadened multi-peak features at larger $U$.  Our results provide reference spectra for this important system, and the techniques can be applied to a variety of other quasi-1D systems.

\section{Acknowledgments}
We thank Mingru Yang, Steven Kivelson, and Doug Scalapino for helpful discussions. The algorithms are implemented using the ITensor\cite{itensor} library and the codes are available under the ITensor/TDVP repository. This work is funded by NSF through Grant DMR-1812558.

\appendix
\section{\label{sec:Aone} Time Evolution methods}

\begin{figure*}[h]
\centering
\includegraphics[width=0.5\linewidth]{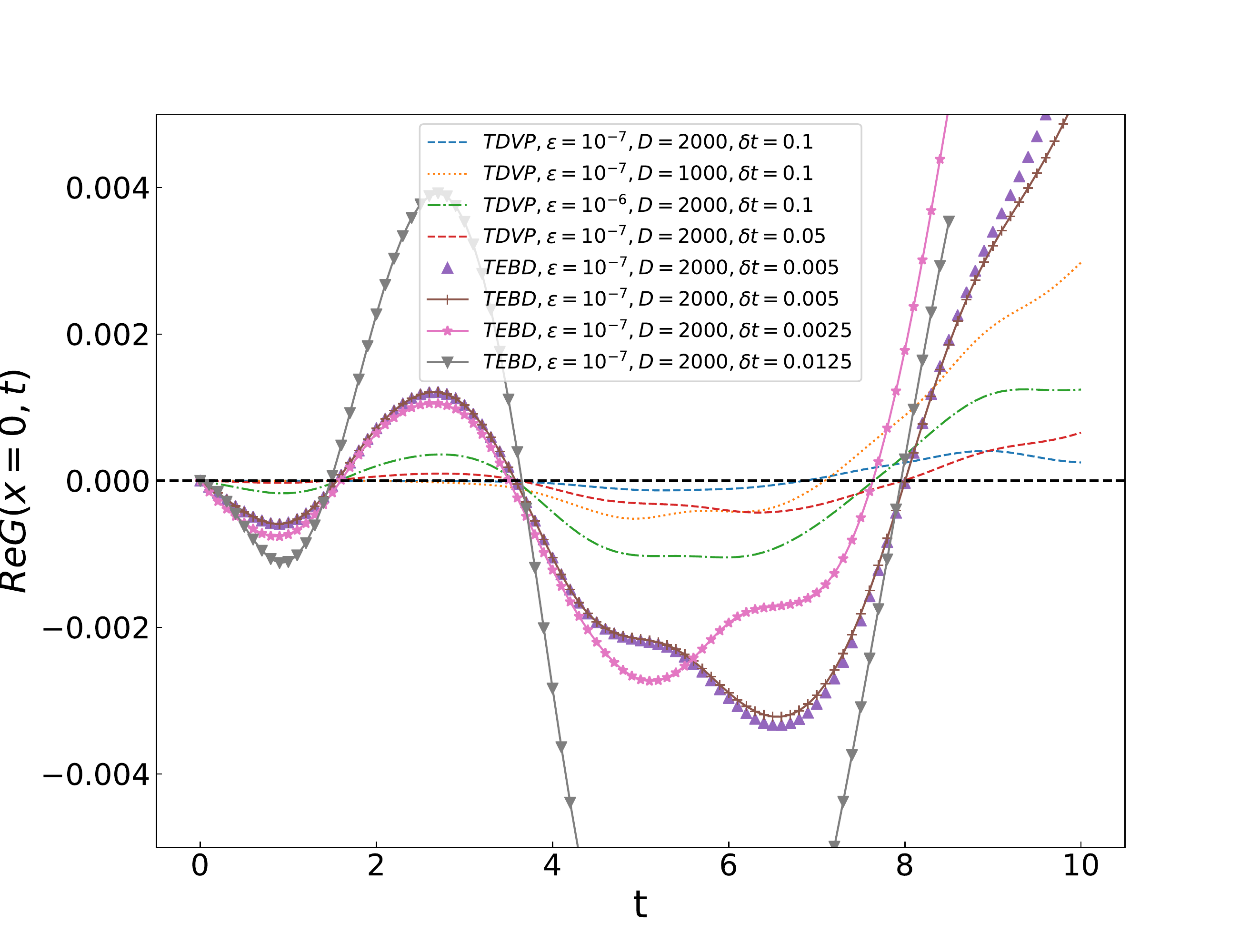}
\includegraphics[width=0.45\linewidth]{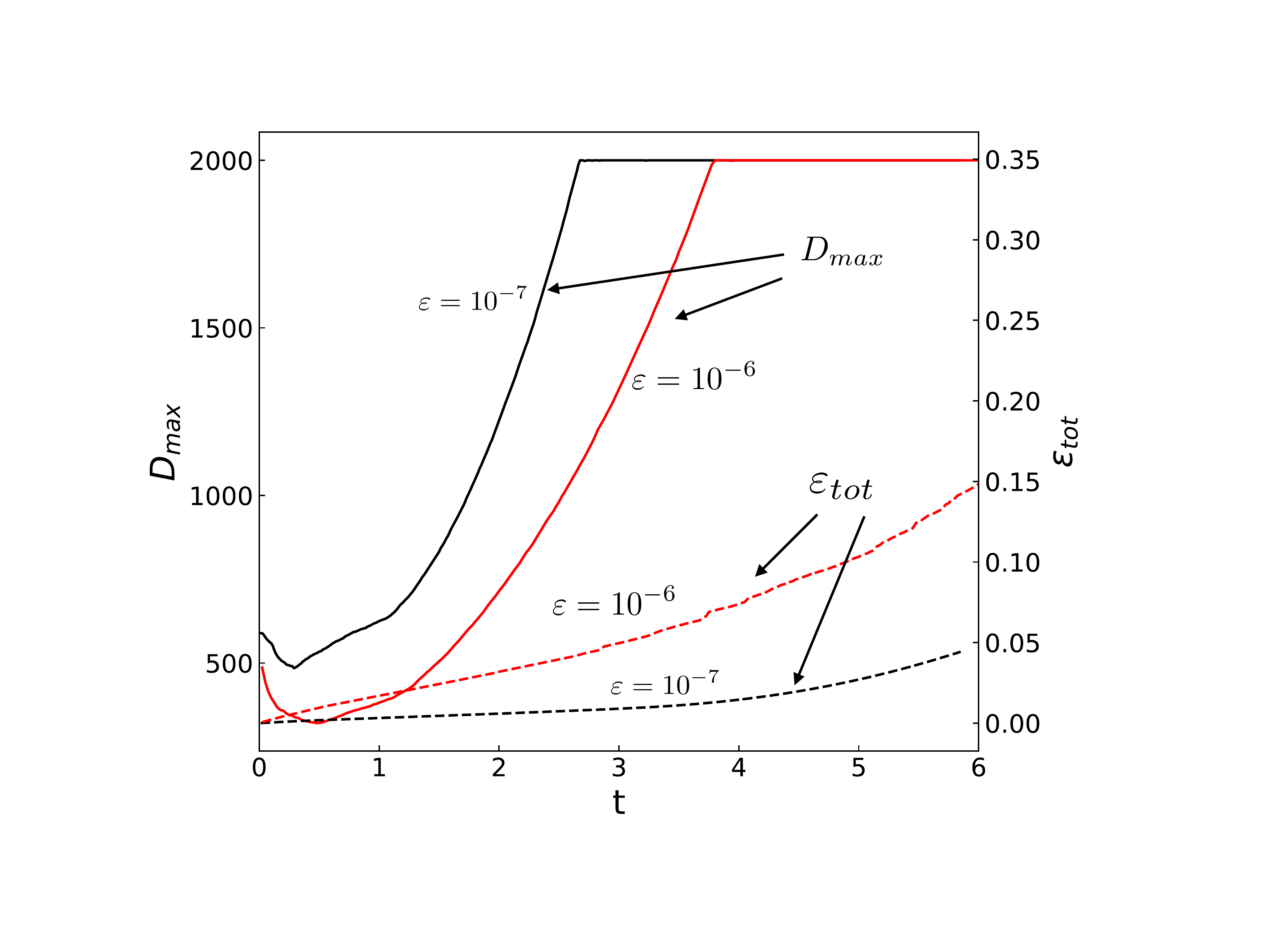}
\caption{(a) shows the difference of Green's function $G(x,t)$ between TDVP and TEBD of different settings. The reference method is TDVP using cutoff $\epsilon=10^{-7}$ and timstep $\delta t = 0.1$. (b) shows the maximum number of states kept $D_{max}$ versus time (upper-left curves) for the $64\times2$ ladder system at cutoff equals $10^{-6}$ and $10^{-7}$ using tDMRG. The simulation adjusted D at each step to try to achieve a total discarded weight of $\epsilon$ for that step, subject to a maximum D of 2000. The TDVP has an similar growth of dimension.}
\label{evolution_method}
\end{figure*}



The implementation of TEBD for the Hubbard ladder model is similar to the tDMRG used in the previous study of $t$-$J$ ladder\cite{White2015}. The tDMRG results were obtained using a Trotter decomposition, applying only nearest-neighbor gates, using a reordering of the sweep path through the lattice to make this possible. The sources of error for tDMRG are finite truncation error and the finite size of the time steps. We tuned the time step to find a time step error that was small enough to have no visible effects. To measure and control the finite truncation errors, we varied the number of states kept (up to D = 2000). 

The main idea of TDVP is to constrain the time evolution to a specific manifold of matrix-product states of a given initial bond dimension. To do so, it projects the action of the Hamiltonian into the tangent space to this manifold and then solves the TDSE solely within the manifold. The source of errors for TDVP is also the finite truncation error and the finite size of the time steps. 

We conducted an experiment with TEBD and TDVP using different settings to find the best efficient setting for the real time evolution on the Hubbard ladder. We used the TDVP with truncation error of $10^{-7}$ and maximum bond dimension D = 3000 as the reference setting for the real-time evolution. The Fig.{\ref{evolution_method}}(a) shows the difference between Green's function $G(x=0,t)$ using different settings and the reference setting. The TDVP's results have a much smaller error compared with TEBD's results. The TEBD is sensitive to the time step. By comparison, The TDVP is not sensitive to the time step and the TDVP using D = 2000 is very close to the TDVP using D = 3000 which has the highest accuracy, indicating an accurate setting used in the production experiments. Both of these two methods have fast growth of the Bond Dimension. The Fig.{\ref{evolution_method}}(b) shows the growth of the Bond Dimension and the analysis of the error. Thus the truncation error is mainly decided by the D.


\bibliography{apssamp}

\end{document}